\documentclass{ws-ijbc}
\usepackage{ws-rotating}     
\usepackage{graphicx}
\usepackage{epstopdf}
\begin{document}

\catchline{}{}{}{}{} 

\markboth{Julia Cantis\'{a}n, Mattia Coccolo, Jes\'{u}s M. Seoane, and Miguel A.F.
	Sanju\'{a}n}{Delay-induced resonance in the time-delayed Duffing oscillator}

\title{Delay-Induced Resonance in the Time-Delayed Duffing Oscillator}

\author{Julia Cantis\'{a}n, Mattia Coccolo, Jes\'{u}s M. Seoane, Miguel A.F. Sanju\'{a}n}

\address{Nonlinear Dynamics, Chaos and Complex Systems Group, Departamento de F\'{i}sica \\ Universidad Rey Juan Carlos, Tulip\'{a}n s/n, 28933 M\'{o}stoles, Madrid, Spain}

\maketitle

\begin{history}
\accepted{1 august 2019}
\end{history}
\begin{abstract}
The phenomenon of delay-induced resonance implies that in a nonlinear system a time-delay term may be used as an effective enhancer of the oscillations caused by an external forcing maintaining the same frequency. This is possible for the parameters for which the time-delay induces sustained oscillations. Here, we study this type of resonance in the overdamped and underdamped time-delayed Duffing oscillators, and we explore some new features. One of them is the conjugate phenomenon: the oscillations caused by the time-delay may be enhanced by means of the forcing without modifying their frequency. The resonance takes place when the frequency of the oscillations induced by the time-delay matches the ones caused by the forcing and vice versa. This is an interesting result as the nature of both perturbations is different. Even for the parameters for which the time-delay does not induce sustained oscillations, we show that a resonance may appear following a different mechanism.
\end{abstract}


\section{Introduction}
The phenomenon of resonance has always been a topic of interest in science and engineering \cite{Rajasekar2016a} because it enables the enhancement of the output signal of an oscillator by means of different types of perturbations. We could classify a resonance for nonlinear systems depending on the nature of the perturbation as: stochastic resonance (SR) \cite{Gammaitoni1998}, when the perturbation is a noise signal; vibrational resonance (VR) \cite{Landa2000}, when it is a high-frequency periodic forcing; and chaotic resonance (CR) \cite{Zambrano2007}, when it is a chaotic signal.

Vibrational resonance and stochastic resonance have also been studied in systems with time-delay \cite{Jeevarathinam2011,Kim1999, Yang2010}. These kind of systems have gained popularity in the scientific community as the time-delay accounts for the finite propagation time, and affects a wide variety of physical, engineering and biological processes. A relevant characteristics of a time-delayed system is that its future evolution depends not only on its present state, but also on a previous state of its history, that we denote with the time-delay $ \tau $. This implies that initial conditions for $t = 0$ are no longer enough to define a particular solution. On the contrary, a history function, i.e., a set of initial conditions in the continuous time interval $ [-\tau, 0] $, is necessary. This is the reason why these systems are said to evolve in an infinite-dimensional phase space \cite{Farmer1982}. Among the branches of science where a time-delay is present, we may consider neural networks since the speed of information transfer in the axons and dendrites is finite \cite{Popovych2011}, population dynamics due to the gestation and maturation times \cite{Kuang1993a,Liu2016}, meteorology due to the transport times of mass or energy from one location of the globe to another \cite{Keane2017}, laser arrays because of the speed of light  \cite{Soriano2013}, and car following models for traffic flow simulation \cite{Orosz2010}. It also plays a vital role in electronics since the speed of modern data processing does not allow to neglect finite propagation times  \cite{Just2010a}.

Concerning the study of the resonance in time-delayed systems, there is some preliminary literature  that analyzes the possibility to enhance the system's response solely by means of a time-delay, excluding perturbations such as a noise, chaotic signals or periodic forcings. We mention here some of these cases. The question of the need of a high-frequency forcing when a time-delay is present was already posed by LV et al. [~\citeyear{Lv2015}]. The effect of the time-delay in the periodically driven damped Duffing oscillator was analyzed in ~\cite{RAVICHANDRAN2012} using a perturbation theory. A new resonance for the time-delayed overdamped Duffing oscillator in the presence of a Bogdanov-Takens bifurcation leading to the phenomenon called Bogdanov-Takens resonance was analyzed in~\cite{Coccolo2018}. 

Furthermore, a time-delay together with a nonlinear term is presented in~\cite{Yang2015} as an effective mechanism to enhance a weak input periodic forcing without the aid of a high-frequency forcing or a noise signal. This is possible because the time-delay induces oscillations in the system of frequency $ \omega_{n}(\tau) $ for certain values of $ \tau $. When this frequency equals the frequency of the oscillator, a phenomenon named as delay-induced resonance arises. 

However, some features concerning this phenomenon remain open: the effect of the amplitude of the time-delay term, the dynamics when the time-delay does not induce oscillations or the possibility to induce a resonance by means of the forcing in presence of the time-delay, among others. Indeed, this phenomenon was only studied for a time-delayed overdamped oscillator with a cubic term but not for other systems.  

In this paper, we aim to broaden the current knowledge on the delay-induced resonance through the study of the periodically driven Duffing oscillator with a time-delay. In particular, we analyze the overdamped and underdamped cases and we will cover the topics previously mentioned. For the overdamped oscillator, we focus on the parameter space and we carry out analytically and numerically a bifurcation analysis of the unforced system in order to determine the parameter values for which the time-delay induces oscillations. For these parameter values, we analyze not only the effect of $ \tau $ (as in~\cite{Yang2015}), but also the effect of the amplitude of the time-delay term, $ \gamma $. We show that a time-delay term with parameters $ (\gamma, \tau) $ may enhance the oscillations produced by an external forcing without the aid of any other kind of perturbation. Furthermore, we study the conjugate phenomenon: the possibility of enhancing the oscillations induced by the time-delay by means of the periodic forcing. Secondly, we show that the signal enhancement is possible even for the parameters for which the time-delay does not induce oscillations. For the underdamped oscillator, a similar analysis is carried out and the phenomenon of delay-induced resonance also arises displaying some differences. 

This paper is organized as follows: Sec.~\ref{Soverdamped} describes the dynamics of the overdamped system and it is divided into three subsections covering the bifurcation analysis, delay-induced resonance for the parameters for which the time-delay induces sustained oscillations and for the parameters for which it induces decaying oscillations. In both cases, the effect of the time-delay and the forcing are considered. Section~\ref{Sunderdamped} describes the underdamped case and it is divided into two subsections covering the numerical bifurcation analysis and the study of resonance for a specific range of parameter values. Again, the effect of the time-delay and the forcing are considered separately. Discussions and conclusions are presented in Sec.~\ref{Sdiscussion}.

\section{Overdamped and time-delayed Duffing oscillator} \label{Soverdamped}

The overdamped oscillator is the limit for which an oscillator $m \ddot{x} + b \dot{x}=F(x) $ with a large frictional force compared to its inertial term ($ b \dot{x} >> m \ddot{x} $) can be reduced to $ b \dot{x}=F(x) $ \cite{Strogatz1994}. In our case, the overdamped system with time-delay and a forcing reads as follows
\begin{equation}
\dot{x} - x +  x^{3} -\gamma x(t-\tau)=g \cos{\Omega t},
\label{overdamped_general}
\end{equation}
where $ \tau $ is the time-delay, $ \gamma $ is the amplitude of the time-delay term and $g \cos{\Omega t}  $ is a weak forcing that produces the oscillations that we aim to enhance.

\subsection{Bifurcation analysis}
We start with the study of the dynamics of the unforced system
\begin{equation}
\dot{x} - x +  x^{3} - \gamma x (t-\tau)=0,
\label{overdamped}
\end{equation}
which is used to model the widely known atmospheric phenomenon of El Ni$\bar{n}$o Southern Oscillation (ENSO) where the variable $ x $ represents the sea surface temperature anomaly \cite{Suarez1988}. 

Without the time-delay term, $ \gamma x (t-\tau) $, Eq.~\ref{overdamped} would reduce to a first order ODE with two symmetric stable fixed points $ x= \pm 1 $ in addition to an unstable fixed point $ x=0 $. This dynamical behavior corresponds to the flow on a line which implies that the system is not able to oscillate under any circumstances. However, the time-delay makes the system infinite-dimensional allowing the possibility of oscillations for certain parameter values $ (\gamma, \tau) $. In other words, the origin of the oscillatory dynamics of Eq.~\ref{overdamped} is the time-delay term. 

A linear stability analysis is performed around the origin, $ x(t)=0 $, in order to determine the parameter region of stability $ (\gamma, \tau) $ for this solution. Linearizing Eq.~\ref{overdamped} near $ x(t)=0 $ gives us
\begin{equation}  
\dot{x}=x(t)+ \gamma x(t-\tau).
\end{equation}
Then, we calculate the characteristic equation which is found setting $ x(t)=Q e^{\lambda t} $, where $ \lambda $ is a complex parameter
\begin{equation}
\lambda=1+\gamma e^{-\lambda \tau}.
\label{eq_caracteristica}
\end{equation}

If Re$(\lambda)<0 $, the fixed point is stable, that is, a trajectory that starts at the origin would remain there and a trajectory that starts elsewhere would decay to the origin after a transient oscillatory regime. When the real part of the eigenvalue changes from being negative to being positive, the fixed point looses stability and the system is said to cross a Hopf bifurcation. This means that the origin is no longer stable and a trajectory starting near the origin would move away in the form of an outward spiral in phase space. As we can appreciate in Eq.~\ref{eq_caracteristica}, the values of $ \gamma $ and $ \tau $ affect the eigenvalue, $ \lambda $,  thus they affect the stability of the fixed point. To find the condition for the Hopf bifurcation, we set $ \lambda= i \omega $ which accounts for the change of sign in Re$(\lambda)$

\begin{equation}
\pm i\omega=1+\gamma e^{\pm i\omega \tau} =1+\gamma (\cos{\omega \tau} \pm i\sin{\omega \tau}).
\end{equation}

Separating the real and imaginary parts, we have
\begin{subequations}
	\begin{align}
	1=-\gamma \cos{\omega \tau}\\
	\omega= -\gamma \sin{\omega \tau}.
	\end{align}
\end{subequations}

After some algebra we finally get the Hopf bifurcation curve given by
\begin{equation}
\tau= \frac{\arccos(-1/\gamma)}{\sqrt{\gamma^{2}-1}},
\label{curve}
\end{equation}
which accounts for the values of $ ( \gamma, \tau) $ for which the origin looses stability.

Another bifurcation takes place when $ \lambda=0 $. This substitution in Eq.~\ref{eq_caracteristica} leads to $\gamma=-1$, which defines a pitchfork bifurcation as the system presents a single zero root. At the point $ (\gamma, \tau)=(-1,1) $, the Hopf and pitchfork bifurcation curves cross, leading to a Bogdanov-Takens bifurcation as Eq.~\ref{eq_caracteristica} presents a double zero root. A detailed analysis of the system's behavior when $ \gamma>-1 $ can be found in~\cite{Redmond2002}. 

\begin{figure}
	\begin{center}
		\includegraphics[width=0.75\textwidth ]{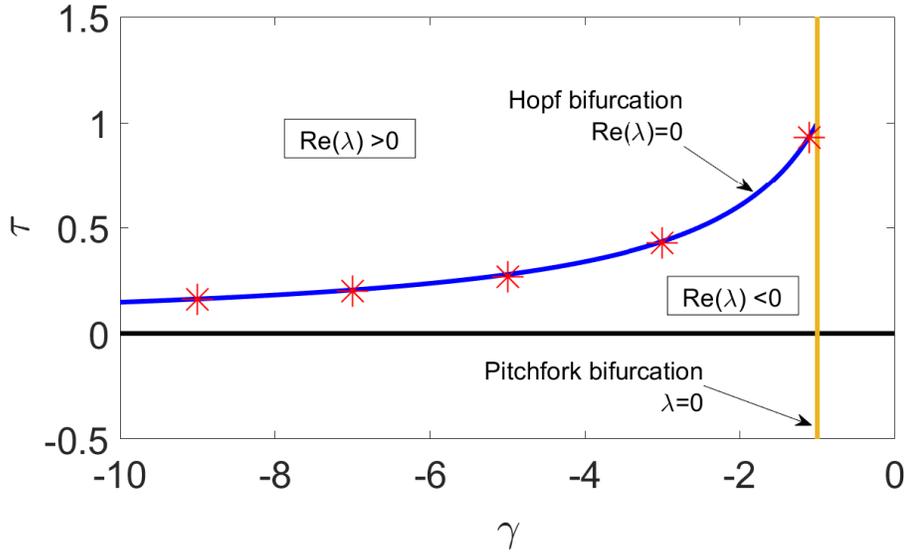}
	\end{center}	
	\caption{\textbf{ Stability diagram in the parameter space $ ( \gamma, \tau) $ for the origin $ x(t)=0 $.} The vertical line corresponds to the the pitchfork bifurcation which marks a change in the number of fixed points, three to the right of the bifurcation line and one to the left. The curve corresponds to the Hopf bifurcation that implies a change in the stability of the fixed point. For values of $ (\gamma, \tau) $ below the Hopf bifurcation, trajectories present decaying oscillations. For values above the Hopf bifurcation, trajectories near the origin diverge in the form of an outward spiral. The points on the curve correspond to numerically calculated points of the Hopf bifurcation curve.}
	\label{Stability_diagram}		
\end{figure}

Both the Hopf and the pitchfork bifurcations are depicted in Fig.~\ref{Stability_diagram}. The parameter space is divided in different regions depending on the system's behavior near the origin. The vertical line corresponding to the pitchfork bifurcation outlines the system's transition from one fixed point on the left of the bifurcation to three fixed points in the region to the right. For a fixed $ \gamma<-1 $, we call $ \tau_{c} $ the corresponding time-delay crossing the Hopf bifurcation curve. A trajectory starting nearby the origin $x(t)=0$ would draw an inward spiral for a value of $ \tau<\tau_c $ and an outward spiral for $ \tau>\tau_c $. 

Numerically calculated points of the Hopf bifurcation curve appear too. These points, which correspond to the stars along the Hopf bifurcation curve in Fig.~\ref{Stability_diagram}, have been calculated by fixing $ \gamma $ and selecting the $ \tau $ for which the oscillations start in phase space. It can be seen that they match the theoretically calculated values by Eq.~\ref{curve}. The history function used was $ u_{0}=0.01 $ so that the trajectory starts nearby the origin $x(t)=0$. Every numerical integration in this paper was performed using the method of steps to reduce our DDE to a sequence of ODEs that are solved by a fixed-step Runge-Kutta algorithm (Bogacki–Shampine $3/2$ method) \cite{Bogacki1989}.

Linear stability analysis only provides information about the system's behavior near the origin. When the effect of the nonlinear term $ x^{3} $ is taken into account, it can be observed that for $ \tau > \tau_c $ a trajectory near the origin diverges as an outward spiral as predicted from linear stability analysis, and also that a limit cycle arises when it is sufficiently far from the origin. This means that values of $ (\gamma,  \tau) $ above the Hopf bifurcation generate sustained oscillations for the unforced system (Eq.~\ref{overdamped}). 

\begin{figure}
	\begin{center}
	\includegraphics[width=0.75\textwidth]{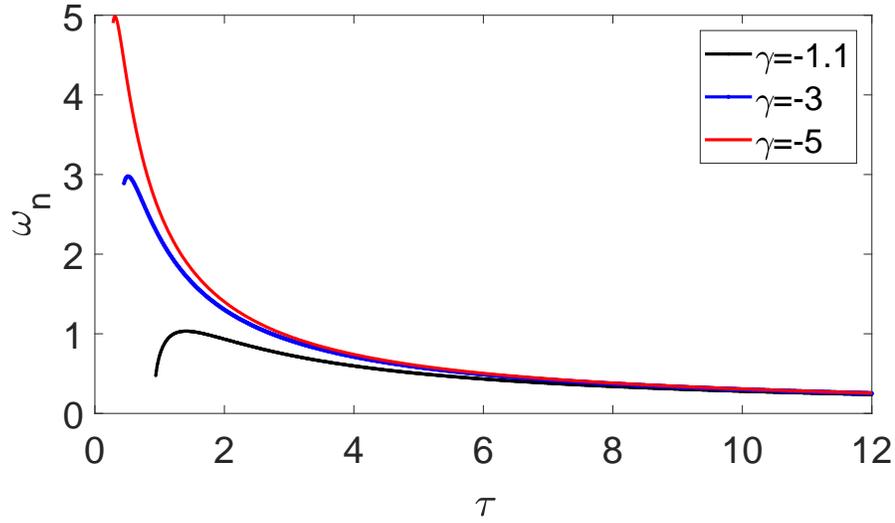} 
	\end{center}
	\caption{ \textbf{Natural frequency, $ \omega_{n} $, induced by the time-delay for $ \tau> \tau_{c}$} . The frequency reaches a maximum for small values of $ \tau $ while for larger values the frequency decays showing the same behavior independently of $ \gamma $.}
	\label{wn_overdamped}
\end{figure}

Here, we study the dynamics above ($ \tau > \tau_c $) and below  ($ \tau < \tau_c $) the Hopf bifurcation. For that purpose, we fixed $ \gamma=-3 $ and through Eq.~\ref{curve} we get the critical value $ \tau_{c} $=0.435 from which our nonlinear system (Eq.~\ref{overdamped}) oscillates. The frequency of these oscillations, which we call the natural frequency, was numerically calculated for different values of $ \tau>0.435 $ and also for some other values of $ \gamma $ (see Fig.~\ref{wn_overdamped}). If the derivative of our variable depends on a state far away in time, that is, $ \tau $ sufficiently large, the frequency induced decays showing the same behavior independently of the amplitude of the time-delay term.  

\begin{figure}
\begin{center}
	\includegraphics[width=0.75\textwidth]{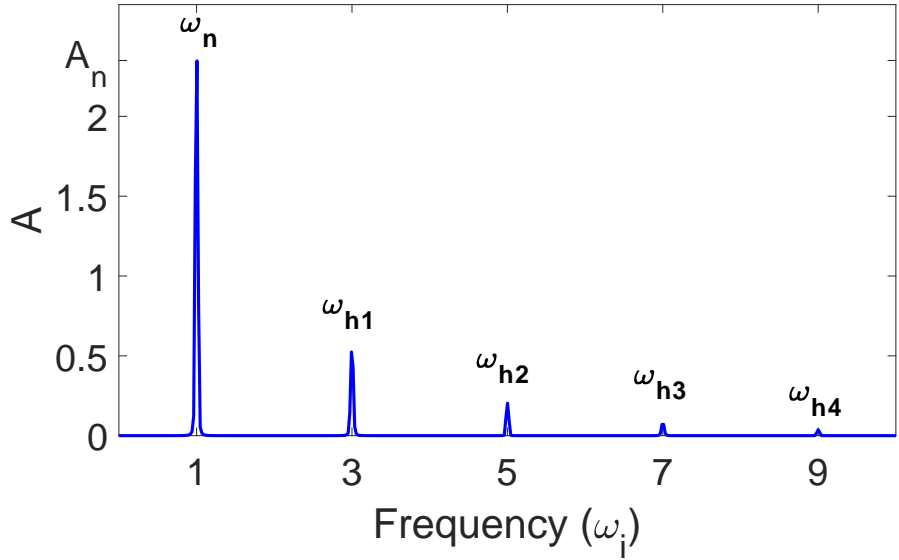} 
\end{center}

	\caption{\textbf{Frequency spectrum of the oscillations induced by the time-delay}. The time-delay term with the parameters $ (\gamma=-3 $, $ \tau=2.73) $ induces multiple frequencies $ \omega_{i} $ in Eq.~\ref{overdamped}. The main peak appears for a frequency that we refer to as $ \omega_{n} $ and we refer to its amplitude as $ A_{n} $. Also, the frequencies $ \omega_{hi} $ with $ i=1, 2, 3... $ correspond to its harmonics.}
	\label{FFT_sinforz}
\end{figure}

Numerically, the frequency is calculated by using the Fourier transform (Fast Fourier Transform algorithm from MatLab that is based on a library called FFTW \cite{Frigo2005}) of the stationary time series for each $ \tau $. Figure~\ref{FFT_sinforz} shows the spectrum for $ \gamma=-3 $, where the first peak corresponds to what we have called the natural frequency. We refer to its amplitude as $ A_{n} $ and $ \omega_{hi} $ with $ i=1,2,3... $ to the frequency of its harmonics, as shown in Fig.~\ref{FFT_sinforz}.  

\subsection{Delay-induced resonance for $ \tau>\tau_{c} $}

In this subsection, we study the resonance for the system described by Eq.~\ref{overdamped_general} for values of $ \tau>\tau_{c} $. For these values, the time-delay induces oscillations of frequency $ \omega_{n} $ together with its harmonics. This implies that the response of the system contains a set of frequency components
\begin{equation}
x(t)=\sum_{i=1}^{n}Q(\omega_{i}) \cos(\omega_{i}t+\phi_{i}), 
\label{response}
\end{equation}
where $ \omega_{i} $ are the frequencies contained in the response such as $ \omega_{n} $ in Fig.~\ref{FFT_sinforz} and $ \phi_{i} $ are arbitrary phases.

If we are interested in the amplitude corresponding to a particular frequency, we may calculate the Q factor which is mathematically defined as

\begin{equation}
Q(\omega_{i})=\sqrt{Q^{2}_{sin}(\omega_{i})+Q^{2}_{cos}(\omega_{i})},
\end{equation}
where $  Q_{sin}(\omega_{i}) $ and $ Q_{cos}(\omega_{i}) $ are
\begin{equation}
Q_{sin}(\omega_{i})=\frac{2}{mT} \int_{0}^{mT}x(t) \sin(\omega_{i}t)dt \\
\hspace{1cm} Q_{cos}(\omega_{i})=\frac{2}{mT} \int_{0}^{mT}x(t) \cos(\omega_{i}t)dt .
\end{equation}

Therefore, the magnitude $ Q(\Omega) $ is a measure of the amplitude of the solution at the frequency $ \Omega $ and $ Q(\omega_{n}) $ is a measure of the amplitude of the solution at $ \omega_{n} $. Numerically, the Q factor is calculated by selecting the amplitude of the peak in the frequency spectrum that corresponds to the frequency we are interested in. 

First, we explore the effect of the time-delay in the magnitude $ Q(\Omega) $ as we aim to enhance the oscillations of the forcing. Secondly, we explore the effect of the forcing in the magnitude $ Q(\omega_{n}) $ as we aim to enhance the oscillations induced by the time-delay.

\subsubsection{Effect of the time-delay}

To explore the effect of the time-delay in Eq.~\ref{overdamped_general}, we initially consider our system as
\begin{equation}
\dot{x} - x +  x^{3} =0.1 \cos{t}.
\label{sindelay}
\end{equation}

\begin{figure} 
	\begin{center}
	\includegraphics[width=0.75\textwidth]{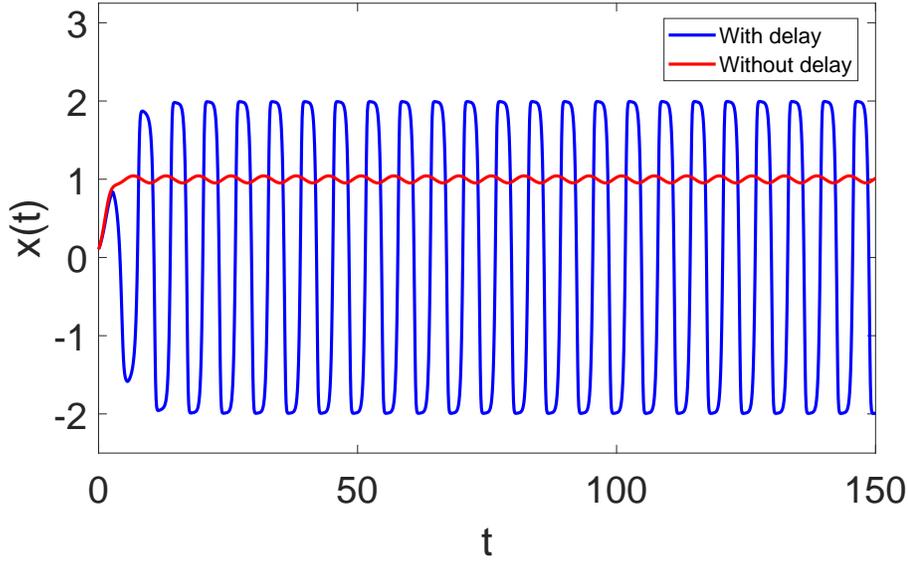}
	\end{center}
	\caption{\textbf{Time series with (large amplitude oscillations) and without (small amplitude oscillations) the time-delay for the forced and overdamped oscillator: $ \dot{x} - x +  x^{3} = 0.1 \cos{t} $}. When a time-delay term such as $ -3 x(t-2.73) $ that induces oscillations of $ \omega_{n}=\Omega $ is considered, the oscillations produced by the forcing are enhanced maintaining the same frequency.}
	\label{fig4}
\end{figure}

The forcing leads to small sustained oscillations of frequency $ \Omega $ around one of the fixed points $ x= \pm 1 $ depending on the initial condition (see Fig.~\ref{fig4}). There are many ways to enhance the system's response, for instance, by adding a second forcing with frequency $ \Omega'>>\Omega $ which is a method based on the vibrational resonance phenomenon. However, we approach the problem using the properties of the time-delay, adding a term like $ -\gamma x (t-\tau) $.

In absence of the forcing, the time-delay induces oscillations with a frequency $ \omega_{n} $, which depends on the parameters $ (\gamma, \tau) $ as shown in Fig.~\ref{wn_overdamped}. Thus, we fixed $ \gamma $ and studied the resonance appearance when  $ \tau $ is varied. However fixing $ \tau $ and varying $ \gamma $ is also possible the appearance of the resonance, what is discussed at the end of this subsection.

\begin{figure}
\begin{center}
	\includegraphics[width=0.75\textwidth]{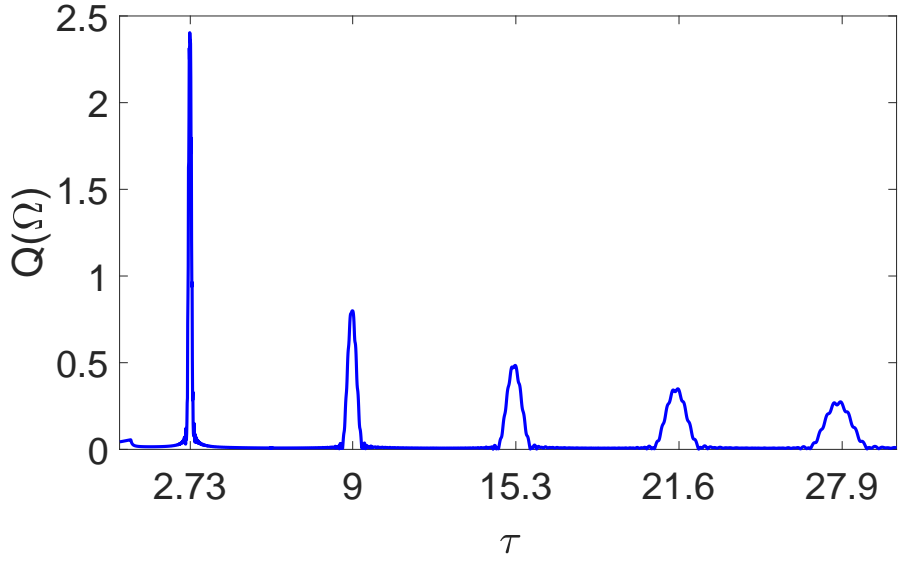} 
\end{center}

	\caption{\textbf{Resonance in the Q factor for the value of $ \tau $ for which $ \omega_{n}= \Omega $}. When a feedback time-delay term such as $ 3x(t-\tau) $ is added to $ \dot{x} - x +  x^{3} = 0.1 \cos{t} $, it can be seen that there is an optimum value of $ \tau $, that is $ \tau=2.73 $, for which the amplitude at the forcing frequency is greatly enhanced. This value corresponds to the $ \tau $ for which the time-delay induces oscillations of $ \omega_{n}= \Omega=1 $. Other local maxima appear for the values of $ \tau $ for which $ \omega_{hi}=\Omega $.}
	\centering
	\label{Q_tau}
\end{figure}

In Fig.~\ref{Q_tau} the resonant behavior is displayed for a time-delay with a fixed amplitude $ \gamma=-3 $ and a first peak appears for $ \tau=2.73 $. For this value, the output signal is enhanced as the condition $ Q(\Omega)>g $  holds \cite{Yang2015}. Also, it is important to remark that these values $ (\gamma=-3, \tau=2.73) $ correspond to a time-delay term that induces oscillations in the unforced system with a natural frequency $ \omega_{n}=\Omega $. What means that when the time-delay induces oscillations of frequency equal to the forcing frequency, then the Q factor is enhanced. Figure~\ref{fig4} shows how the system oscillates around the origin with a larger amplitude when the time-delay is considered. Moreover, the time-delayed system maintains the same frequency $ \Omega$, because the Q factor at the frequency $ \Omega $ for $ \tau=2.73 $ has a maximum. For other parameter values of $ (\gamma, \tau) $, oscillations may be enhanced but the frequency would be changed.  

Furthermore, a delay-induced resonance does not only occur for $ \omega_{n}=\Omega $, instead other local maxima are reached periodically, but their amplitudes decrease. In Fig.~\ref{Q_tau} some of these secondary peaks can be appreciated, the first of them at $ \tau=9 $. These peaks appear because the time-delay does not only induce the natural frequency, but also its harmonics as shown in Fig.~\ref{FFT_sinforz} for $ \tau=3 $. For $ \tau=9 $, its first harmonic ($ \omega_{h1} $) corresponds to $ \Omega $ leading to this secondary peak of resonance. For other values of $ \tau $, the frequency of one of the harmonics may equal the frequency of the forcing leading to successive weaker peaks which appear with a period of $ 2\pi/\Omega$. In Table~\ref{tab}, we include some of these values of $ \tau $ and the frequency component that each resonance peak induces.

\begin{table}[h]
	\tbl{\textbf{Values of the time-delay $ \tau $ that induce a frequency component equal to the forcing frequency.} The values of $ \tau $ appear with periodicity $ 2\pi/\Omega$ and they correspond to the resonance peaks in Fig.~\ref{Q_tau}. The first value $ \tau=2.73 $ induces oscillations of natural frequency equal to $ \Omega $, while for the following values of $ \tau $ it is one of their harmonics that appears at a frequency $ \Omega $.\label{tab}}
	{\begin{tabular}{ |c | c | } 
		\hline
		~ ~ Value of $\tau ~ ~$ &  ~~ Frequency Component ~~ \\ 	[6pt]
		\hline
		$~ ~\tau=2.73 ~ ~$ &  $~~ \omega_{n}=\Omega ~~$ \\ 	[6pt]	
		\hline
		$ \hspace{-1em} \tau=9 $ & $ \omega_{h1}=\Omega $  \\ 	[6pt]	
		\hline
		$~ ~ \tau=15.3 ~ ~$ & $ \omega_{h2}=\Omega $ \\ 	[6pt]	
		\hline
		$~ ~ \tau=21.6 ~ ~ $ & $ \omega_{h3}=\Omega $ \\[6pt]
		\hline
	\end{tabular}}. 
\end{table}

For a low-frequency forcing ($ \Omega << 1 $), the value of the time-delay $ \tau $ for which the resonance appears would be independent of $ \gamma $, since the curves in Fig.~\ref{wn_overdamped} converge at small values of $ \omega_{n} $ ($ \omega_{n} << 1 $). 

\begin{figure}
\begin{center}
	\includegraphics[width=0.75\textwidth]{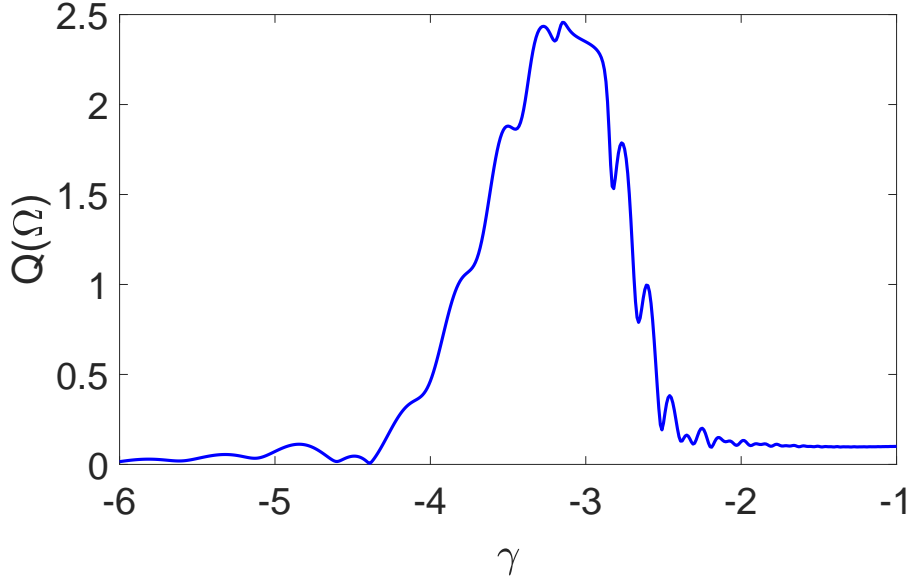} 
\end{center}
	\caption{\textbf{Resonance in the Q factor for the value of $ \gamma $ for which $ \omega_{n}= \Omega $}. When a feedback time-delay term of the form $ -\gamma x(t-2.73) $ is added to $ \dot{x} - x +  x^{3} = 0.1 \cos{t} $, we observe that there is an optimum value of $ \gamma $, that is $ \gamma=-3 $, for which the amplitude at the forcing frequency is greatly enhanced. As it happened with $ \tau $,  this value corresponds to the $ \gamma $ for which the time-delay induces oscillations of $ \omega_{n}= \Omega $. }
	\label{Q_gamma}
\end{figure}

In Fig.~\ref{Q_gamma} the Q factor in function of $ \gamma $ is shown when $ \tau $ is fixed to $2.73$. We recall that the condition to stay on the left of the pitchfork bifurcation (Fig.~\ref{Stability_diagram}) is $ \gamma < -1 $, thus the sweep in $ \gamma $ is restricted. As before, the resonance appears for the $ \gamma $ that induces oscillations of frequency $ \omega_{n}=\Omega $, that is $ \gamma=-3 $. Again, the signal is enhanced at this point since $ Q(\Omega)>g $. The width of the resonance peak is larger in this case as for the fixed $ \tau $ the curves in Fig.~\ref{wn_overdamped} for different $ \gamma $ are very close. This means that a substantial change in $ \gamma $ produces a slight change in $ \omega_{n}$, thus there is an interval of values around $ \gamma=-3 $ that causes the resonance.

A restriction for this type of resonance to appear is that the forcing amplitude has to be small compared to the oscillation induced by the time-delay, otherwise the effect of the time-delay cannot be noticed. 

All in all, when the time-delay induces oscillations of the same forcing frequency, the response amplitude at this frequency reaches a maximum. This result is interesting as we shall remember that for VR the frequencies of both forcings were not equal, one had to be greater than the other.

\subsubsection{Effect of the forcing}
\begin{figure}
\begin{center}
\includegraphics[width=0.75\textwidth]{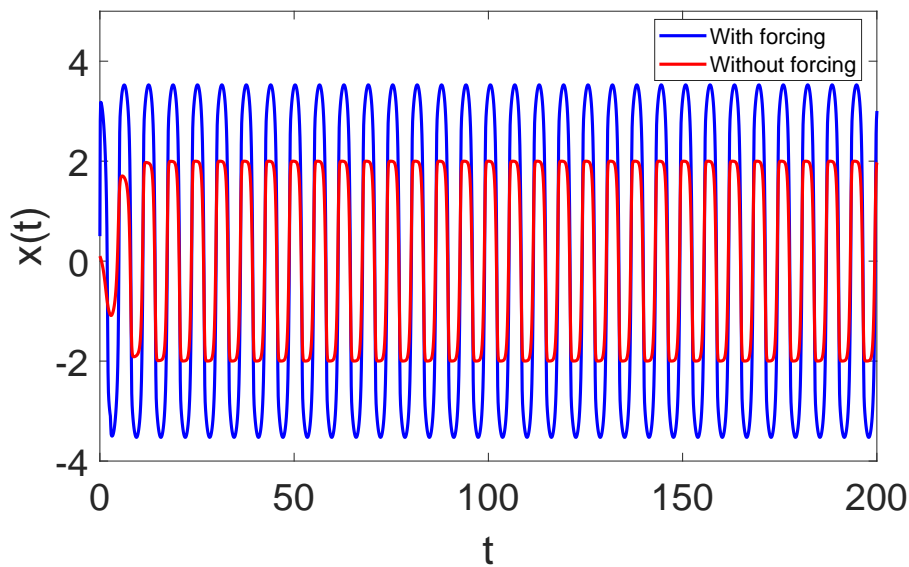} \\ 
\end{center}
	\caption{\textbf{Time series with (large amplitude oscillations) and without (small amplitude oscillations) the forcing for the time-delayed and overdamped oscillator:  $ \dot{x} - x +  x^{3} + 3 x(t-2.73) = 0 $}.  The system without the forcing oscillates due to the time-delay term with frequency $ \omega_{n} $. When a forcing such as $ 30 \cos{\omega_{n} t} $ is added, the amplitude is enhanced maintaining the same frequency.}
	\label{timeseries_conysinforz}
\end{figure}

In this subsection, we aim to enhance the oscillations caused by the time-delay instead of the ones caused by the forcing. The unforced system with time-delay reads
\begin{equation}
\dot{x} - x +  x^{3} + 3 x (t-2.73)=0,
\label{sinforz}
\end{equation}
which oscillates around zero with frequency $ \omega_{n} $ (see Fig.~\ref{timeseries_conysinforz}). Now, a periodic perturbation, $ g \cos{\Omega t} $, is added in order to enhance the response to the frequency $ \omega_{n} $ provoked by the time-delay. The criterion to affirm that the signal caused by the time-delay has been enhanced is that $ Q(\omega_{n})>A_{n} $, where $ A_{n} $ is the amplitude of the unforced system at this frequency, i.e., the first peak in the frequency spectrum (Fig.~\ref{FFT_sinforz}). 

\begin{figure}
\begin{center}
	\includegraphics[width=0.75\textwidth]{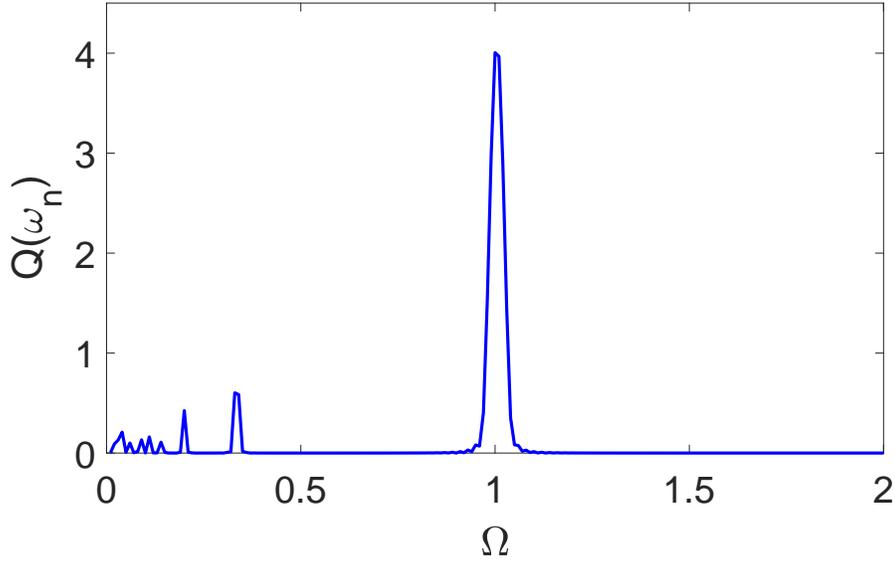} 
\end{center}
\caption{\textbf{Resonance in the Q factor for a forcing with frequency $ \Omega=\omega_{n} $}. When a forcing $ 30 \cos{\Omega t} $ is added to $ \dot{x} - x +  x^{3} + 3 x (t-2.73)=0 $, it can be seen how the resonance appears when the forcing frequency $ \Omega $ equals the frequency of the oscillations induced by the time-delay term which for this case is $ \omega_{n}=1 $. Some minor peaks that do not accomplish the condition $ Q(\omega_{n})>A_{n} $ are present for smaller values of $ \Omega $.}
	\label{Q_omega}
\end{figure}

\begin{figure}
\begin{center}
	\includegraphics[width=0.75\textwidth]{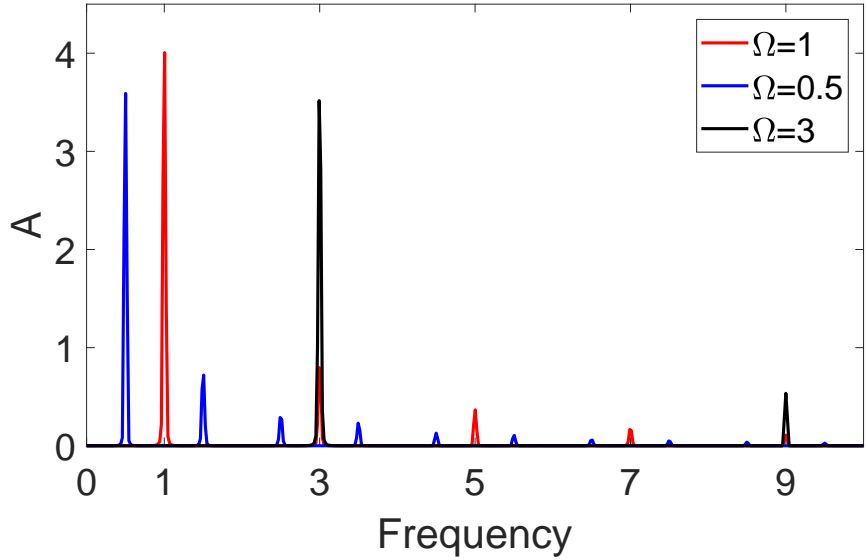} \\ 
\end{center}
\caption{\textbf{Frequency spectrum for $ \dot{x} - x +  x^{3} + 3 x(t-2.73) = 30 \cos{\Omega t} $ with different values of $ \Omega $}. The maximum peak amplitude is achieved when $ \Omega=\omega_{n}=1 $ as the effect of the time-delay and the forcing are added. For other values of $ \Omega $, the system presents a different set of frequencies with slightly smaller amplitudes.}
	\label{FFT_variosomega}
\end{figure}

Figure~\ref{Q_omega} shows that a resonance peak appears in the Q factor when $ \Omega =\omega_{n}$ that fulfills the criterion mentioned above. As for $ (\gamma=-3, \tau=2.73) $, the natural frequency is $ \omega_{n}=1 $, the resonance peak appears for $ \Omega=1 $. This peak grows monotonically with the forcing amplitude $ g $. In Fig.~\ref{timeseries_conysinforz} we show how the oscillations are enhanced when the forcing with $ \Omega=\omega_{n} $ is added, and how the frequency of the oscillations remains the same. For values of $ \Omega \neq \omega_{n}$ the system's response would contain a different set of frequencies, but the response is slightly larger precisely when $ \Omega =\omega_{n}$ because the effect of the time-delay and the forcing are added (see Fig.~\ref{FFT_variosomega}).

A restriction for this type of resonance to appear is that the forcing has to be of greater magnitude than the oscillation caused by the time-delay $ g>A_{n} $, otherwise the effect of the forcing cannot be noticed. 

Comparing Figs.~\ref{Q_tau},~\ref{Q_gamma} and~\ref{Q_omega} we realize that the response of the system is similar in all cases as the resonance takes place when both frequencies have the same value. The equivalence of the effect of a change in $ \Omega $ and a change in $ \omega_{n} $, which is the consequence of a change in $ (\gamma,\tau) $, is not trivial as the nature of both perturbations is different. This fact allows us to talk about conjugate resonances \cite{Blekhman2004}.

\subsection{Delay-induced resonance for $ \tau<\tau_{c} $}

Now, we explore the existence of a resonance in the overdamped and time-delayed Duffing oscillator (Eq.~\ref{overdamped_general}) for $ \tau<\tau_{c} $. In this region of $ \tau $, the dynamics of the unforced system is restricted to decaying oscillations around the origin (without the time-delay, the decay towards the fixed point would be monotonic). However, as we will see, when the time-delay is present a resonance phenomenon appears. Therefore, this makes us consider that in this case the time-delay term plays the role of the enhancing perturbation like the noise in SR or the second periodic forcing in VR. The periodic forcing plays the role of the oscillation inducer. 

First of all, we investigate the effect of the time-delay on the amplitude of the system's response by changing $ \tau $ for values smaller than $ \tau_c $, for a fixed $ \gamma $. Then, we investigate the effect of the frequency and the amplitude of the forcing, for a given time-delay.

\subsubsection{Effect of the time-delay}

For the region $\tau<\tau_{c} $, the sweep in $ \tau $ goes from 0 to $ \tau_{c} $, which can be calculated for a given $ \gamma $ with Eq.~\ref{curve}. In Fig.~\ref{Ahopf}(a) we analyze the peak to peak amplitude response with $ \tau $ for different values of $ g $. We have to keep in mind that there is not a frequency related with the effect of the time-delay on the unforced system for this range of $ \tau $, thus it does not make sense to calculate the Q factor and we use the peak to peak amplitude.

For some values of $ g $, a resonant behavior with $ \tau $ is found (Fig.~\ref{Ahopf}(a)). The value of the time-delay $ \tau_{r} $ for  which the resonance appears, increases with $ g $ leading to obtain, for $ g= 1 $, a monotonic growth of the amplitude for the range of $ \tau  $ considered. A similar restriction for the forcing amplitude was also present in the range $\tau>\tau_{c}$. However, for smaller values of $ g $, the system displays resonant behavior as the amplitude reaches a maximum for $ \tau_{r} $ and then falls again. As it can be seen, for $ \tau_{r} $ the response amplitude is greater than the forcing amplitude, $ g $.

\subsubsection{Effect of the forcing}
Resonant behavior appears as well in Eq.~\ref{overdamped_general} when $ \Omega $ is varied. This phenomenon has also been called \textit{nonlinear resonance in a system with time-delay} as, for example, in~\cite{RAVICHANDRAN2012} where it was explored for the underdamped Duffing oscillator. 

We have chosen the parameters $ (\gamma, \tau) $ so that we lay in the stability region near the Hopf bifurcation, however for values of the parameters further from the Hopf bifurcation the resonance is less intense but is still present. For $ \tau=0.7008 $, the value of $ \gamma $ corresponding to the Hopf bifurcation curve (Eq.~$\ref{curve}$) is $\gamma=-1.65 $. Having this in mind, we choose $ \tau=0.7008 $ and $ \gamma=-1.6 $, which is slightly under the Hopf bifurcation curve in the parameter space. 

\begin{figure}
	\begin{center}	
	\includegraphics[width=8cm]{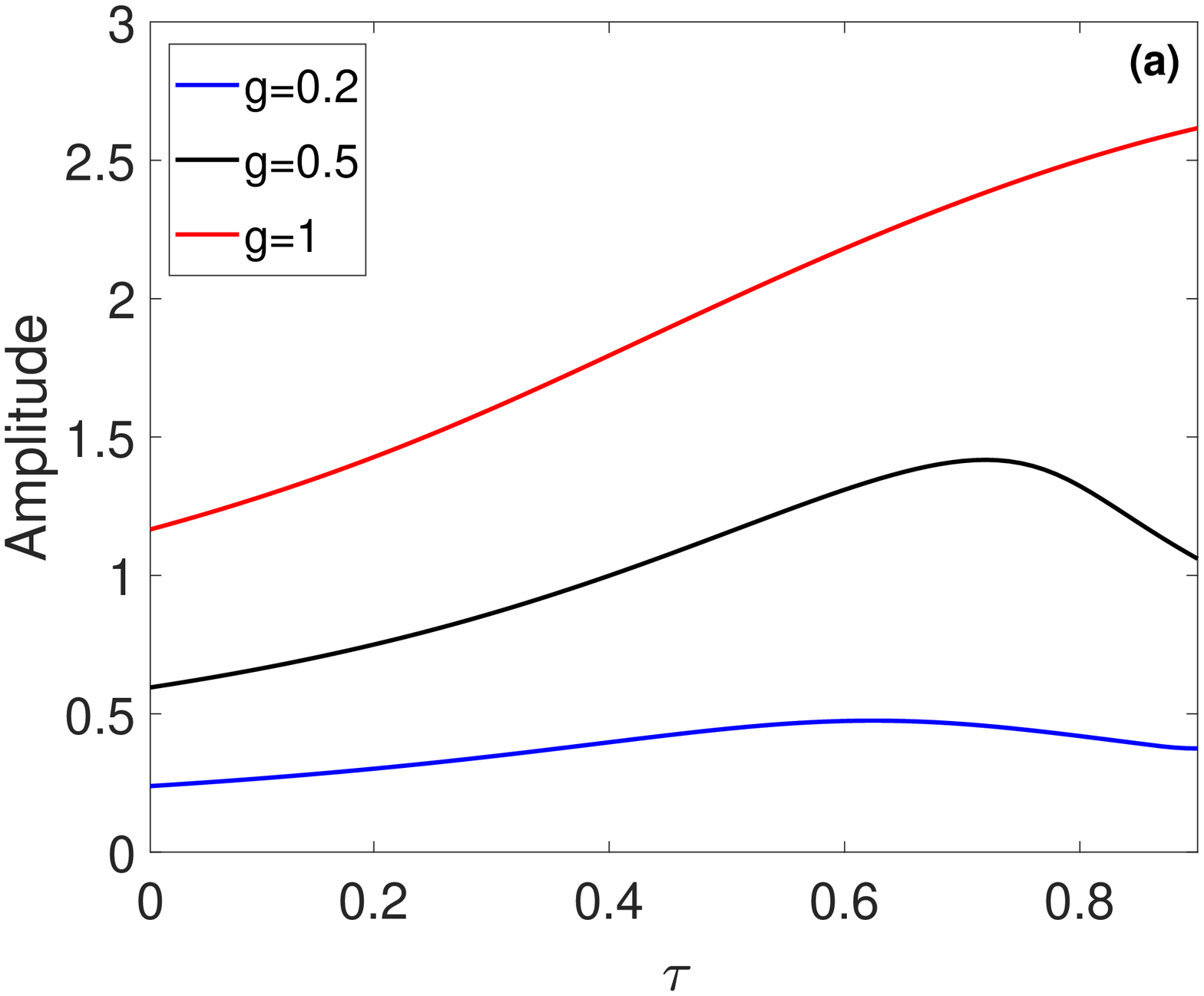}
	\includegraphics[width=8cm]{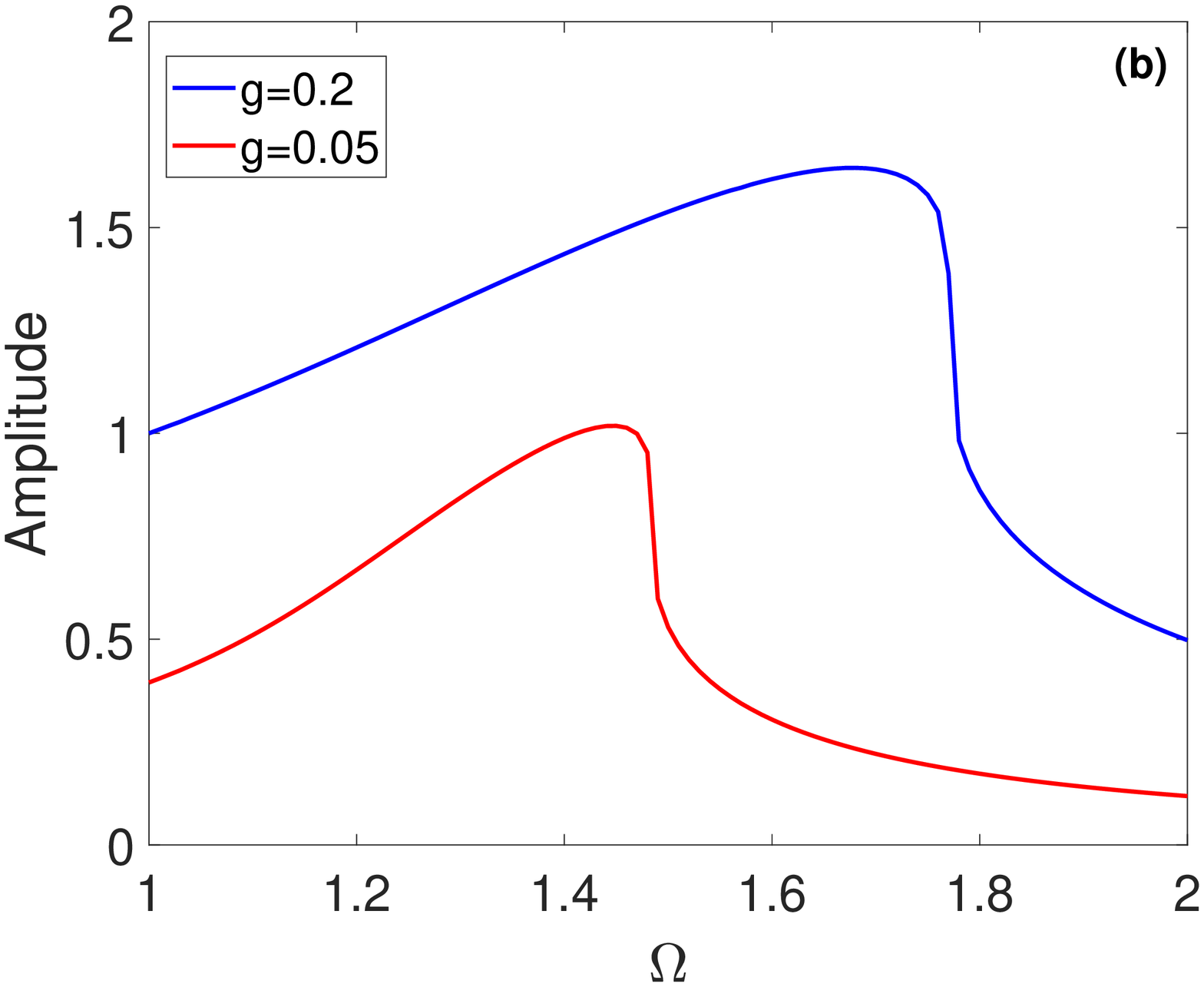}
	\end{center}	
	\caption{ \textbf{Peak to peak amplitude response in function of $ \tau$ and $ \Omega $ for $ \dot{x} - x +  x^{3} -\gamma x(t-\tau)=g \cos{\Omega t} $.}  (a) The panel shows that the resonance appears for a certain value of $ \tau $ that increases with $ g $ leading to monotonic growth for $ g=1 $. The parameter values are $ \gamma=-1.1 $ and $ \Omega=1.69 $. (b) The panel shows that the system also possesses resonant behavior with $ \Omega $, even for very small values of the forcing. The parameter values are $ \tau=0.7008 $ and $ \gamma=-1.6 $. }
	\label{Ahopf}
	
\end{figure}

In Fig.~\ref{Ahopf}(b), we show how the amplitude of the oscillations varies non-monotonically with the forcing frequency, i.e., exhibiting resonant behavior. Also, Fig.~\ref{Ahopf}(b) shows that even for very small values of the forcing such as $ g=0.05 $, we still induce large amplitude oscillations in the system. For higher values of the forcing amplitude, the amplitude of the oscillations grows and the frequency for which the resonance appears, $ \Omega_{r} $, also increases.

\section{Underdamped and time-delayed Duffing oscillator} \label{Sunderdamped}

Once the delay-induced resonance has been analyzed for the overdamped oscillator, we continue with the underdamped system for the parameters considered in \cite{Rajasekar2016a} which make the system bistable
\begin{equation}
\ddot{x} - x + 0.1x^{3} +\gamma x (t-\tau)=g\cos{\Omega t}.
\label{system1}
\end{equation}

The dynamics of the system is richer in this case due to the cooperation of the second derivative, the time-delay and the external forcing. The phenomenon of delay-induced resonance appears as well, although it presents some different characteristics.

\subsection{Bifurcation analysis}

The equation for the unforced underdamped oscillator reads
\begin{equation}
\ddot{x}-x+0.1 x^{3} +\gamma x (t-\tau)=0.
\label{underdamped_free}
\end{equation}

As previously stated, the system is bistable for the chosen parameters, what means that there are three fixed points: one unstable at the origin and two others ($ \pm x^{*} $) corresponding to the bottom of the wells. Unlike in the overdamped oscillator and due to the second derivative, Eq.~\ref{underdamped_free} without the time-delay already presents sustained oscillations that are confined to one of the wells. In the following paragraphs, we explore the effect of the time-delay on these oscillations.

\begin{figure}
\begin{center}
		\includegraphics[width=0.75\textwidth]{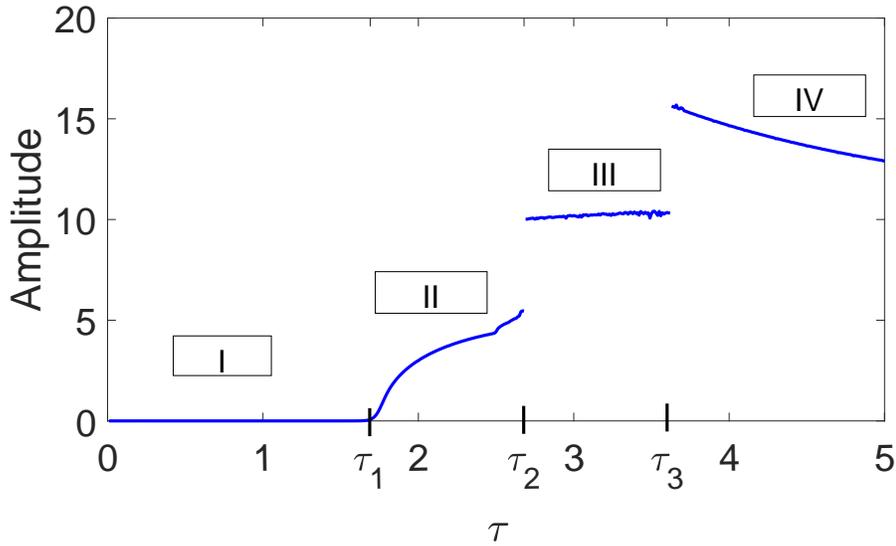} 
\end{center}
\caption{\textbf{Maximum peak to peak amplitude in function of $ \tau $ in Eq.~\ref{underdamped_free}}. The figure shows that four different regions can be distinguished. In the first region (Region I: $ \tau \in (0, \tau_{1}= 1.76)$), the time-delay acts as a damping term for the oscillations caused by the second derivative. In the other regions, the time-delay determines whether the oscillations are confined in one well (Region II: $ \tau \in  [\tau_{1}=1.76, \tau_{2}=2.68)$) or not (Regions III: $ \tau \in [\tau_{2}= 2.68 , \tau_{3}=3.6]$ and IV: $ \tau > 3.6$). The peak to peak amplitude was calculated neglecting the initial transients.}

	\label{amplitude_tau}
\end{figure}

\begin{figure}
	\begin{center}
		\includegraphics[width=7cm]{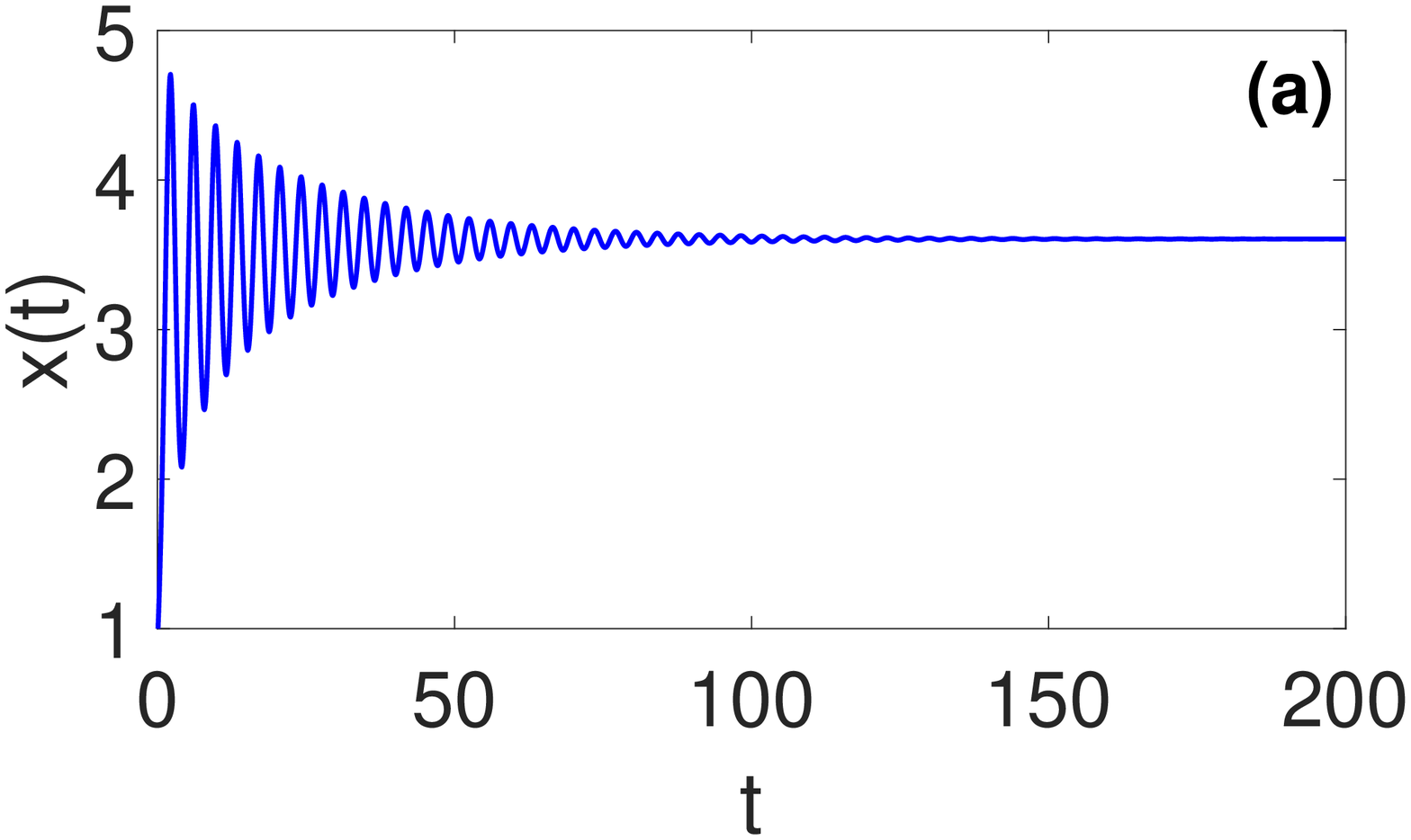}
		\includegraphics[width=7cm]{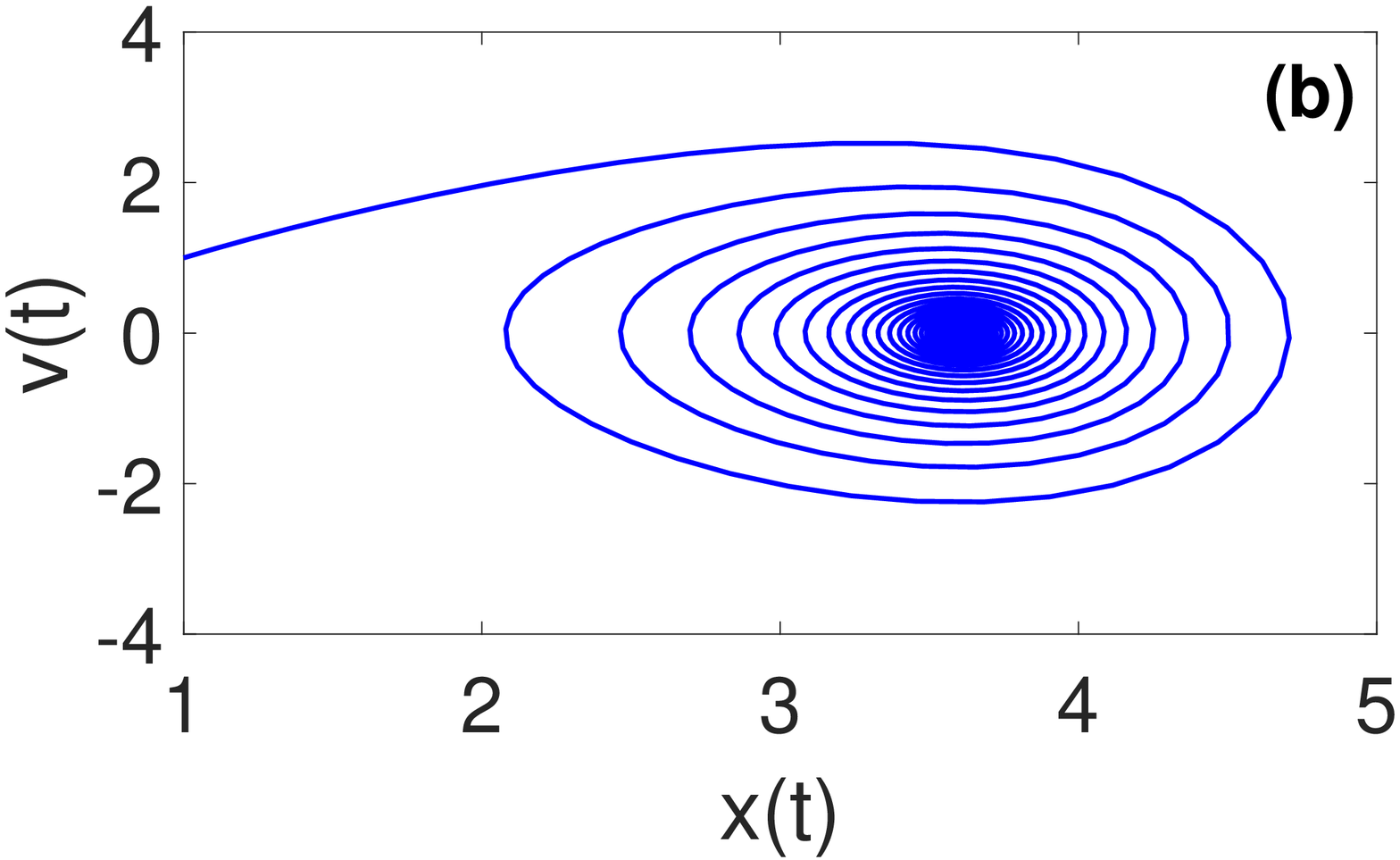} \\
		\includegraphics[width=7cm]{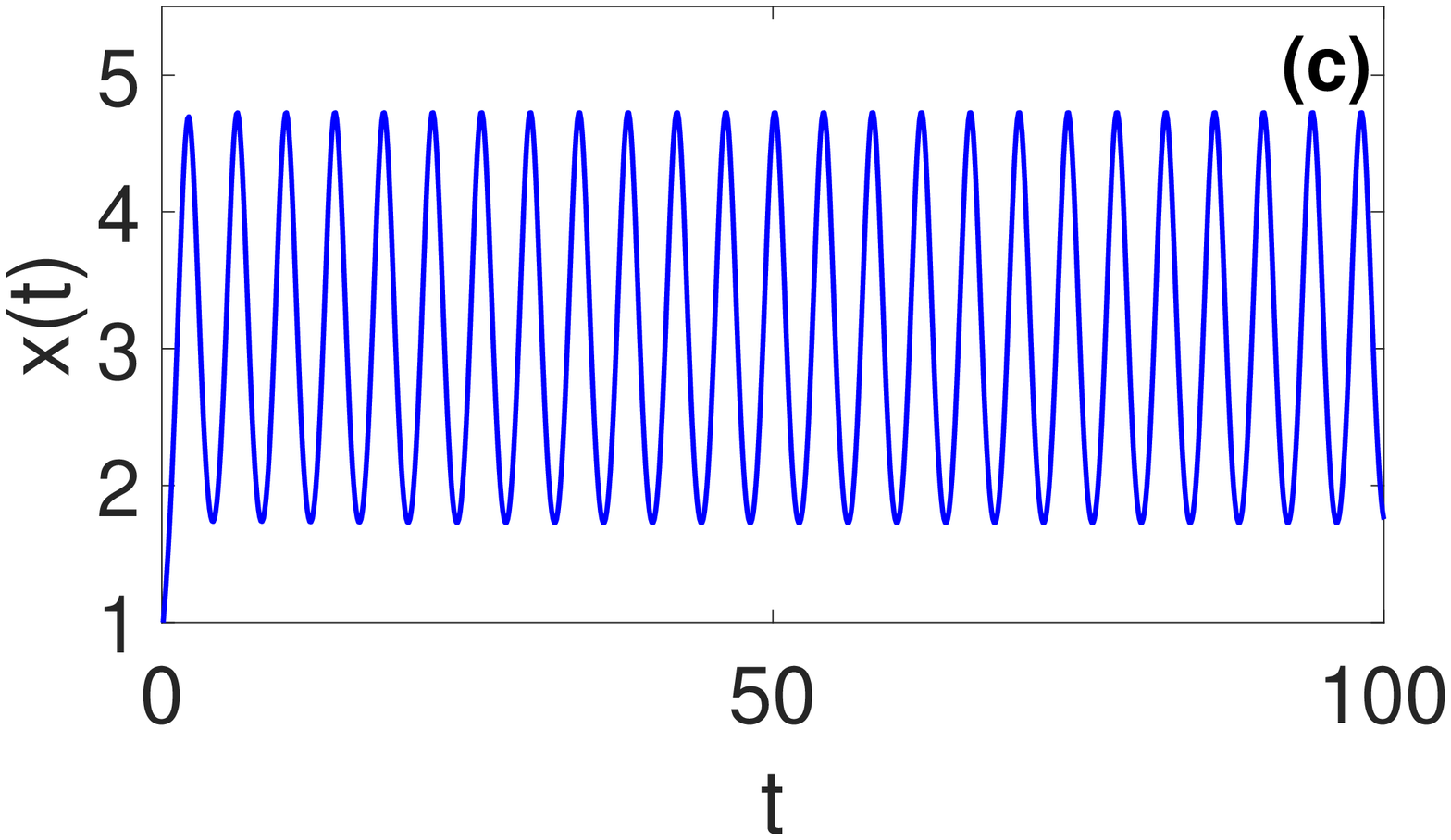}
		\includegraphics[width=7cm]{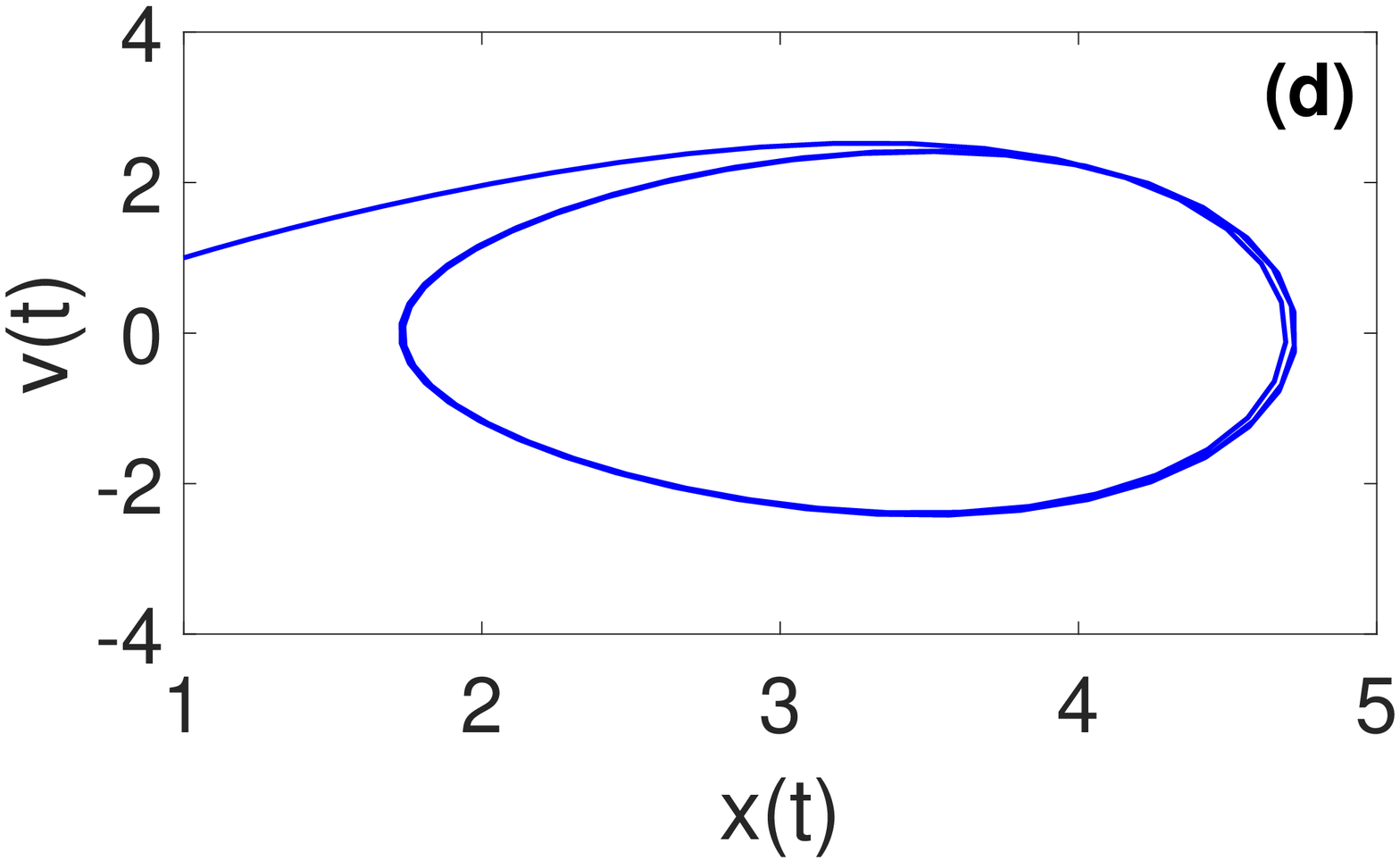}\\
		\includegraphics[width=7cm]{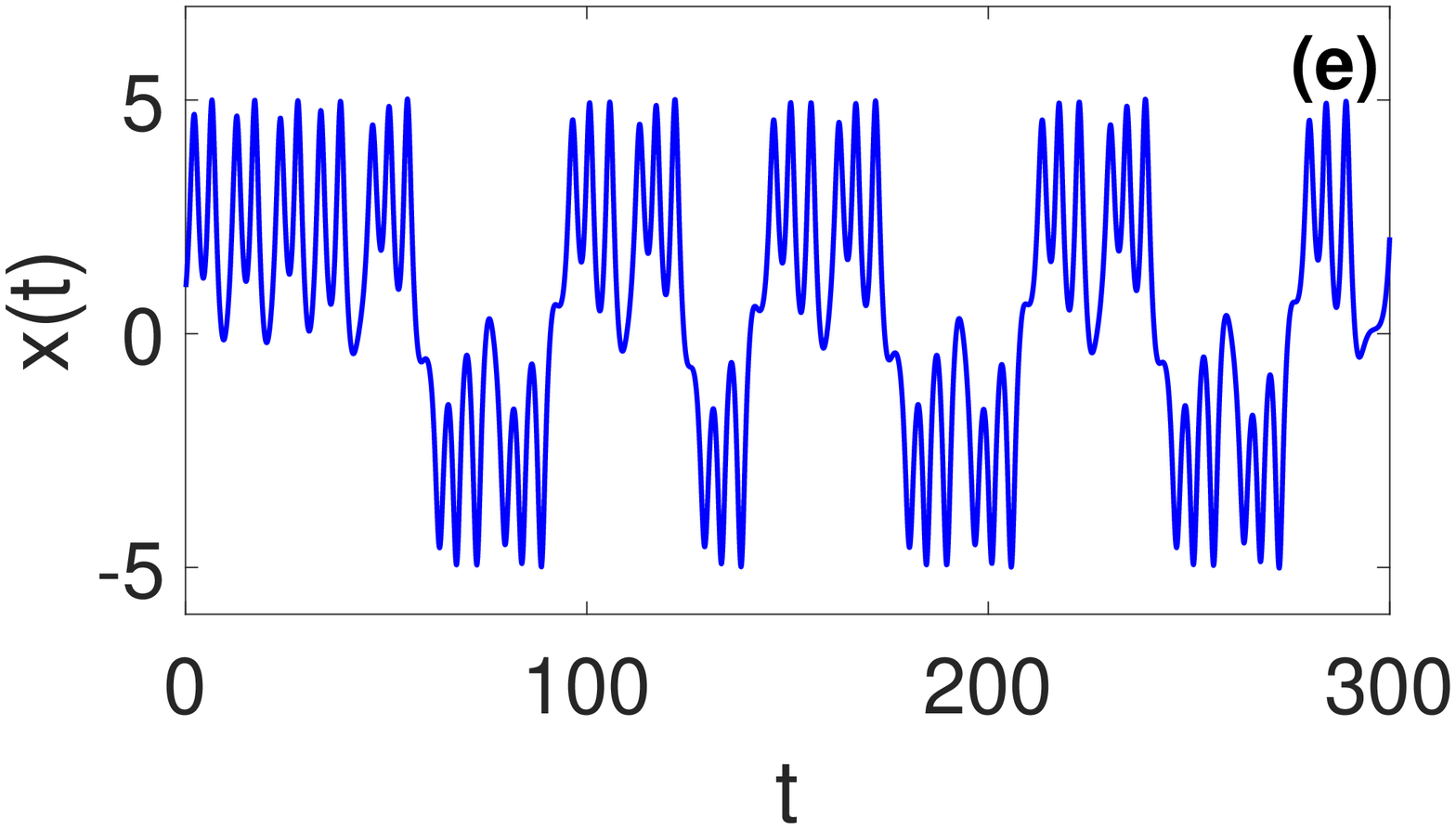} 
		\includegraphics[width=7cm]{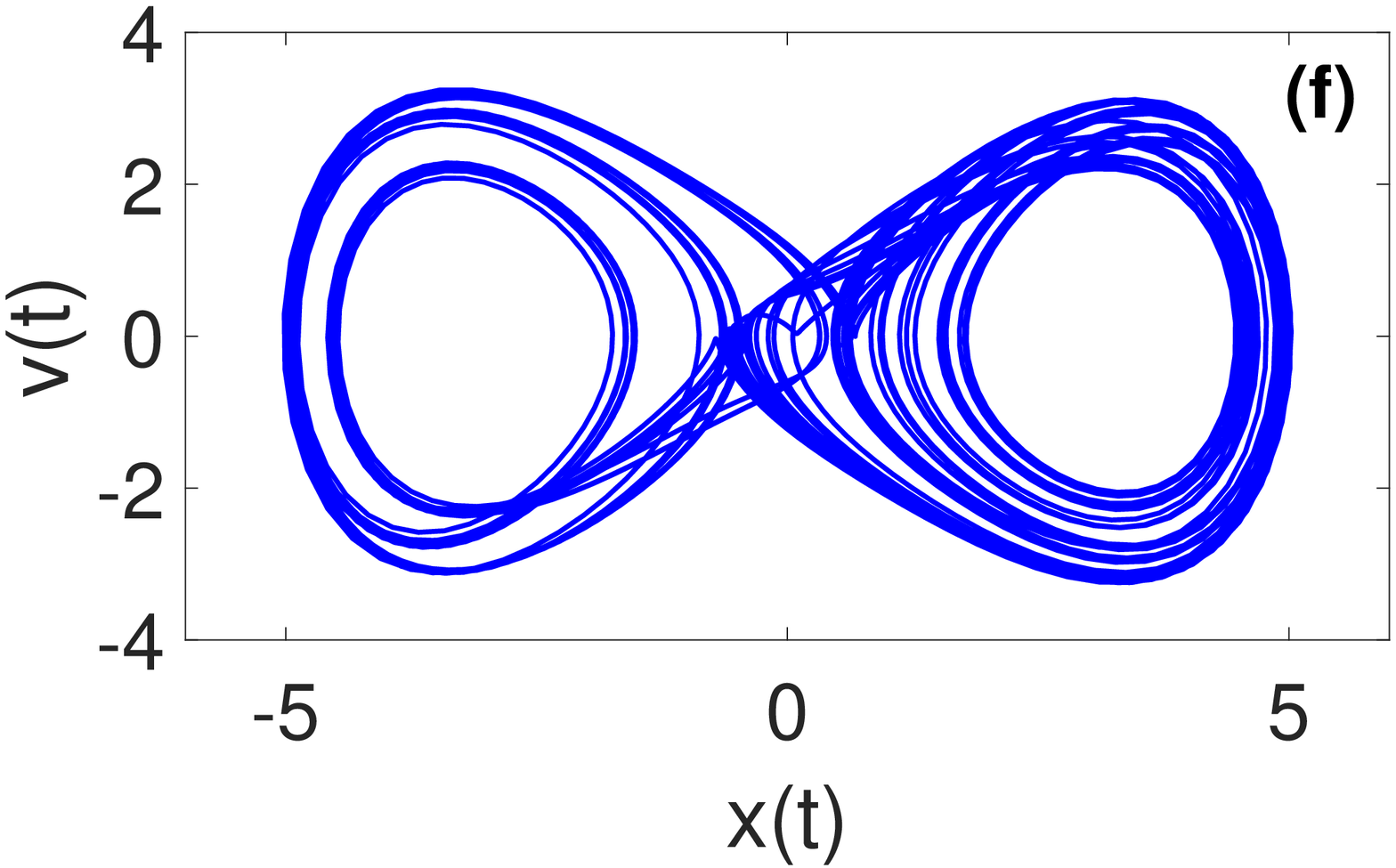} \\
		\includegraphics[width=7cm]{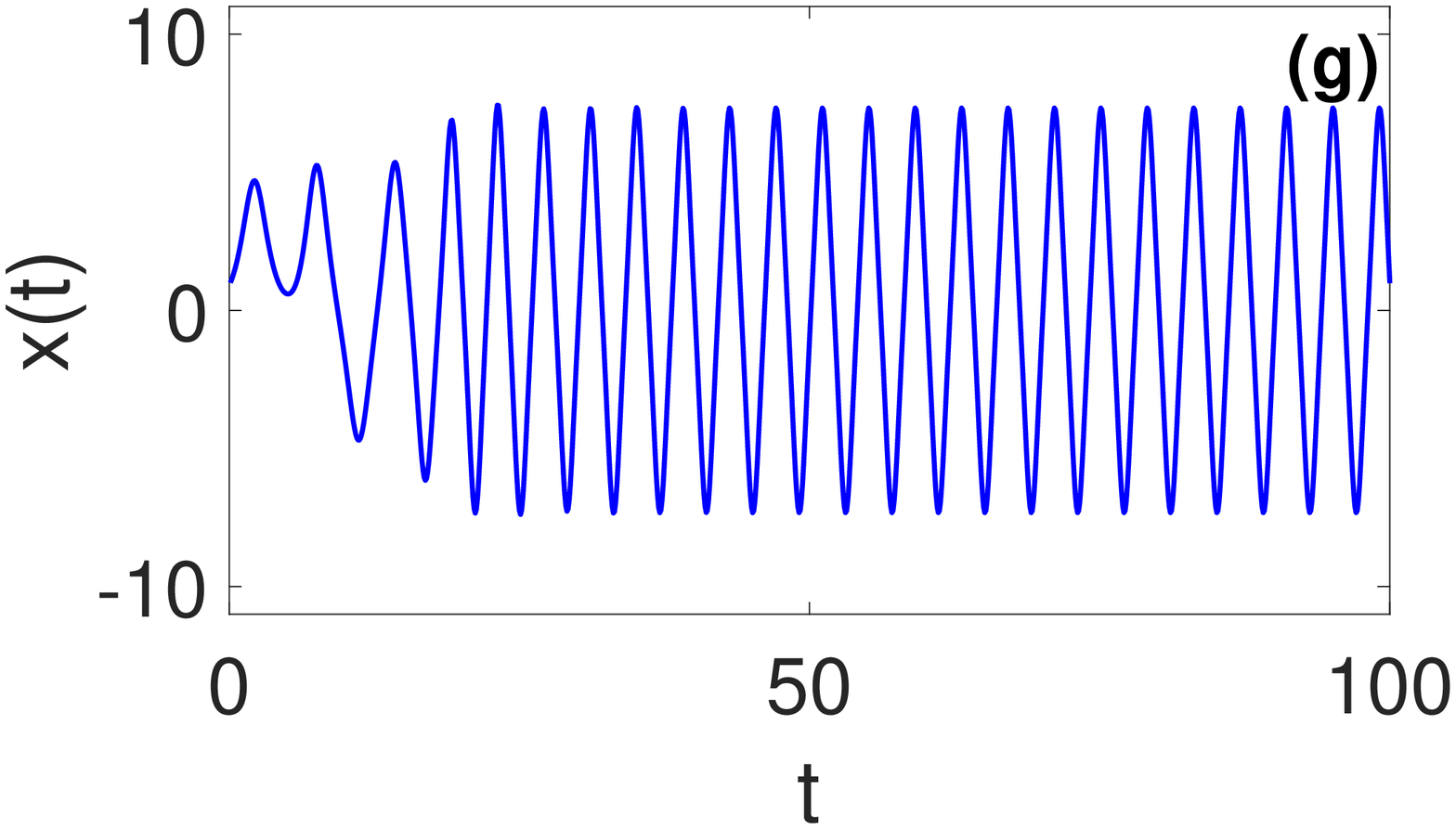}
		\includegraphics[width=7cm]{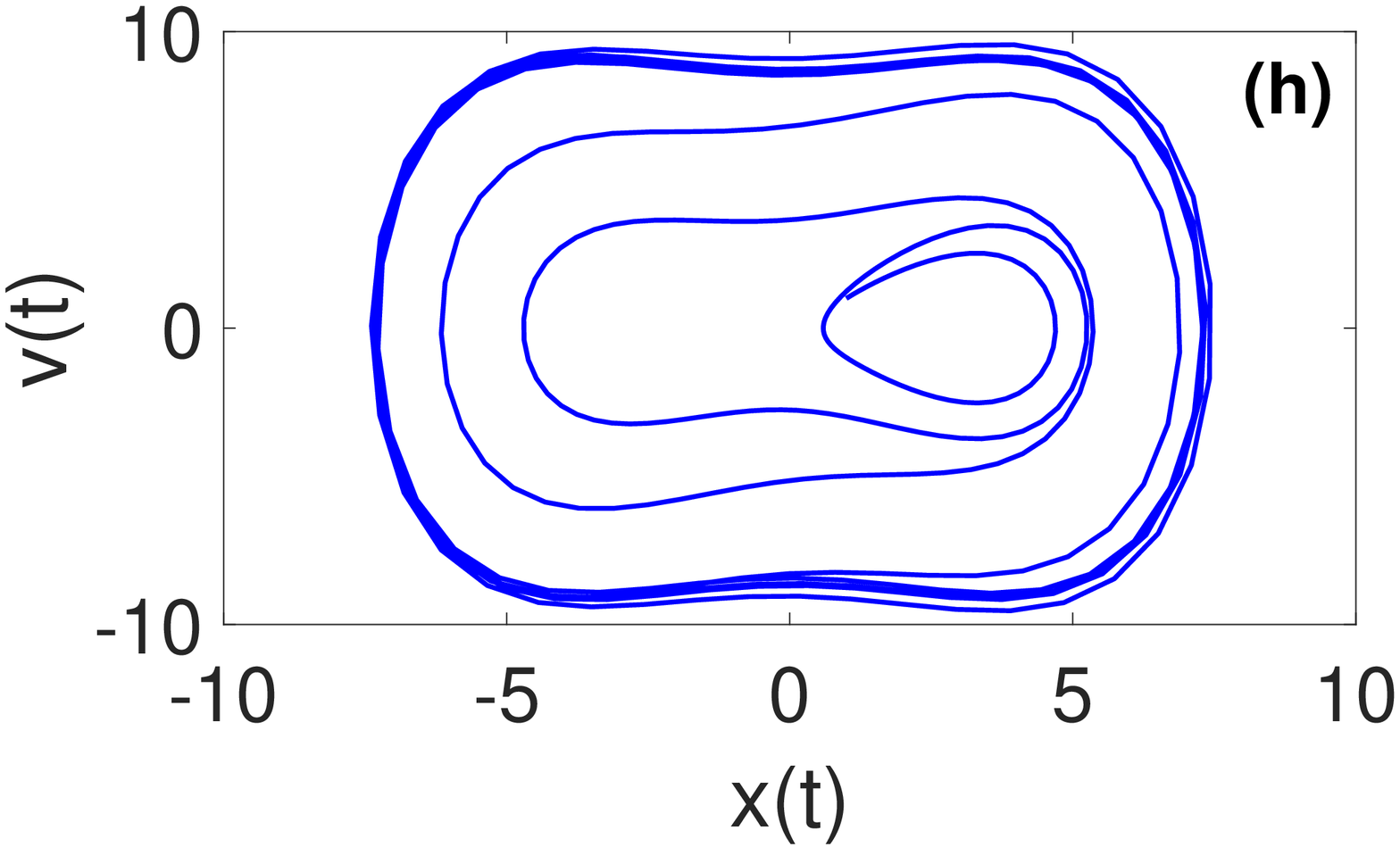} \\
	\end{center}	
	\caption{\textbf{Time series (left column) and phase space (right column) trajectories for values of $ \tau $ corresponding to the four different regions.} (a)-(b): For $ \tau $=1.5 (Region I) oscillations decay with time towards one of the wells. (c)-(d): For $ \tau=2 $ (Region II) oscillations are sustained and bounded in one well. (e)-(f): For $ \tau=2.7$ (Region III), we observe transitions between wells. (g)-(h): Finally, for $ \tau =4$ (Region IV) we are in the last region and the system spans both wells. In these figures, a transient behavior is shown together with its equilibrium state.}
	\label{regions}
\end{figure}

For that purpose and because the dynamics is more complex, we carry out a numerical bifurcation analysis instead of a an analytical one, where we fix $ \gamma=-0.3 $. Then, we study how the dynamics changes with $ \tau $ by plotting the values of the peak to peak amplitude of the time series. By doing so, we can distinguish four different regions (Fig.~\ref{amplitude_tau}) delimited by $ \tau_{1}, \tau_{2}$ and $ \tau_{3} $. The peak to peak amplitude values for each $ \tau  $ were calculated by substracting the minimun of the time series to the maximum. This implies that these values are not necessarily smooth with $ \tau $, specially in the regions where oscillations are not periodic. For simulation purposes a constant history function $ u_{0}=1 $ was used. Also, in Fig.~\ref{regions} we have represented the time series and phase space for four different particular values of $ \tau $ corresponding to the four different regions. 

In the following paragraphs we detail the dynamics of the system in each of the four regions that appear in Fig.~\ref{amplitude_tau}: 

\begin{itemize} 
	\item Region I ($ \tau \in (0, 1.76)$): For the first region, oscillations decay with time until the system reaches one of the fixed points $ \pm x^{*} $. Thus, the steady-state peak to peak amplitude in this region is zero. The time-delay acts as a damping term in the sense that it damps out the oscillations caused by the second derivative \cite{Berezansky2015}. The system falls into the positive well, that is $ x \approx 3.6$ (see Figs.~\ref{regions}(a)-(b)), because the history function is within the basin of attraction for that well. 
	\item Region II ($ \tau \in  [1.76, 2.68)$): Here, the oscillations are sustained and confined to one of the wells. However, the behavior in this region is not homogeneous and we can distinguish two ranges. In the first range, for values of $ \tau \in  [1.76 , 2.5)$, the amplitude of the oscillations is constant and trajectories are periodic (see Figs.~\ref{regions}(c)-(d)). On the other hand, for $ \tau \in [2.5, 2.68)$, trajectories are still confined, but oscillations are not periodic. As already stated, in the latest interval, the maximum peak to peak amplitude values are not smooth with $ \tau $ due to the aperiodicity.
	\item Region III ($ \tau \in [2.68 , 3.6]$): In this region, we see how the amplitude has jumped to a value bigger than the width of the well, what means that trajectories move from one well to another, while without the time-delay, the motion was always confined to one of them. This region is not smooth due to the fact that oscillations are aperiodic. In Figs.~\ref{regions}(e)-(f), we can see how oscillations start in the positive well and after some time the system jumps to the other well where it oscillates aperiodically for some time before jumping to and fro between wells again and again.  
	\item Region IV ($ \tau > 3.6$): In the fourth region, trajectories are no longer confined to neither of the wells and oscillations are periodic. This results in a limit cycle in phase space (Fig.~\ref{regions}(h)) that spans both wells.
\end{itemize}

\begin{figure}
	\centering
	\includegraphics[width=0.75\textwidth]{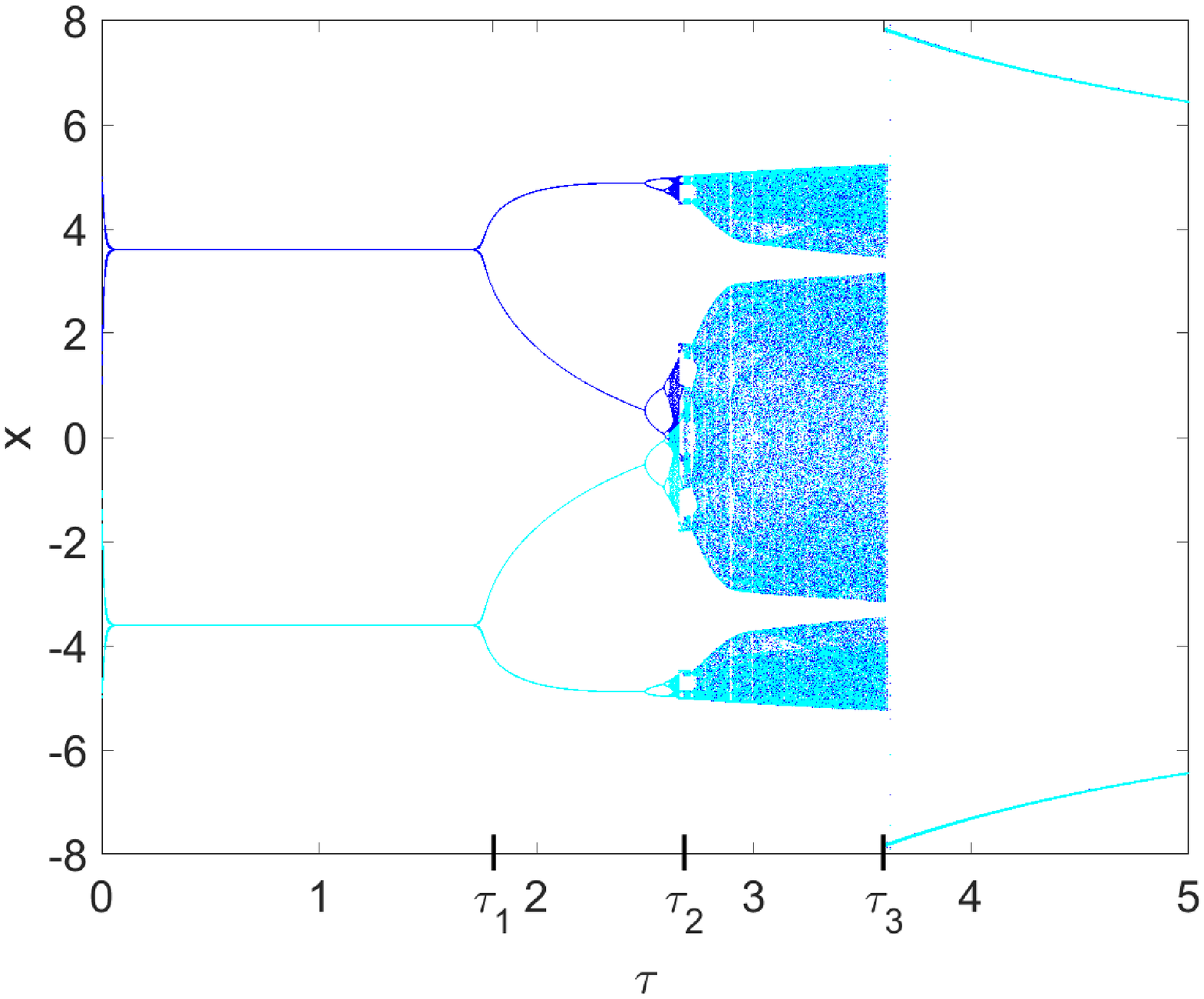} 
	\caption{\textbf{Bifurcation diagram for $ \ddot{x}-x+0.1 x^{3} - 0.3 x (t-\tau)=0 $}. The ticks in the x-axis help us distinguish the four different regions: Region I $[0,\tau_{1}=1.76)$, Region II $[\tau_{1}=1.76, \tau_{2}=2.68 )$, Region III $[\tau_{2}=2.68, \tau_{3}=3.6]$ and Region IV from $ \tau_{3}=3.6 $ on. Two attractors are plotted, the upper one corresponding to the history function $ u_{0}=1 $ and the lower one corresponding to $ u_{0}=-1 $. Chaotic behavior is found at the end of Region II and in Region III and for Regions III and IV both attractors intermingle.}
	\centering
	\label{diagbif}
\end{figure}

To finish the analysis of the system's response to the time-delay, a bifurcation diagram was computed (Fig.~\ref{diagbif}). Two symmetric branches can be seen corresponding to different history functions, each one leading to a different attractor. Several steps were followed to construct the diagram: firstly, it is necessary to extract a map from our continuous system. For this purpose a return map was constructed using the maxima and minima from the time series. We discarded three quarters of these points in order to avoid the transient behavior and plot the last quarter. The same process was repeated for different values of $ \tau $. For the upper branch, the history for the first value of $ \tau $ is a constant function $ u_{0}=1 $, so that, we start in the basin of attraction of the positive well. While for the lower branch we choose $ u_{0}=-1 $, and we start in the basin of attraction of the other well. We used as history functions for the rest of $ \tau $ values the solution of the previous iteration. This way, we make sure that we stay in the same basin of attraction as the parameter is varied \cite{Sprott2011}. Otherwise, the system may jump to another attractor while varying the parameter and we would obtain a meaningless bifurcation diagram. 

The diagram shows the typical period-doubling route to chaos. In Region II, we find two different regimes both confined to one well: for $ \tau \in [1.76, 2.5) $ the period of the system is two, while for $ \tau \in [2.5, 2.68) $, the period starts doubling reaching a chaotic regime with some periodic windows. Furthermore, in Region III,  $ \tau \in [\tau_{2}, \tau_{3}] $, the system that was initially confined to one of the wells starts jumping from one to another as the two chaotic attractors intermingle. For values larger than $ \tau_{3} $, i.e., Region IV, periodicity appears again. 

For the rest of the study of the resonance for the underdamped oscillator, we restrain to values of $ \tau \in [1.76, 2.5) $, the periodic range of Region II. We believe this region is more interesting as oscillations are confined to one well, unlike in Regions III and IV, and with the cooperation of the time-delay and the forcing, these oscillations can be enhanced so that they span both wells. With respect to Region I, the time-delay acts as a damping term so that a delay-induced resonance is not possible for that range of $ \tau $.

\begin{figure}
	\centering
	\includegraphics[width=0.75\textwidth]{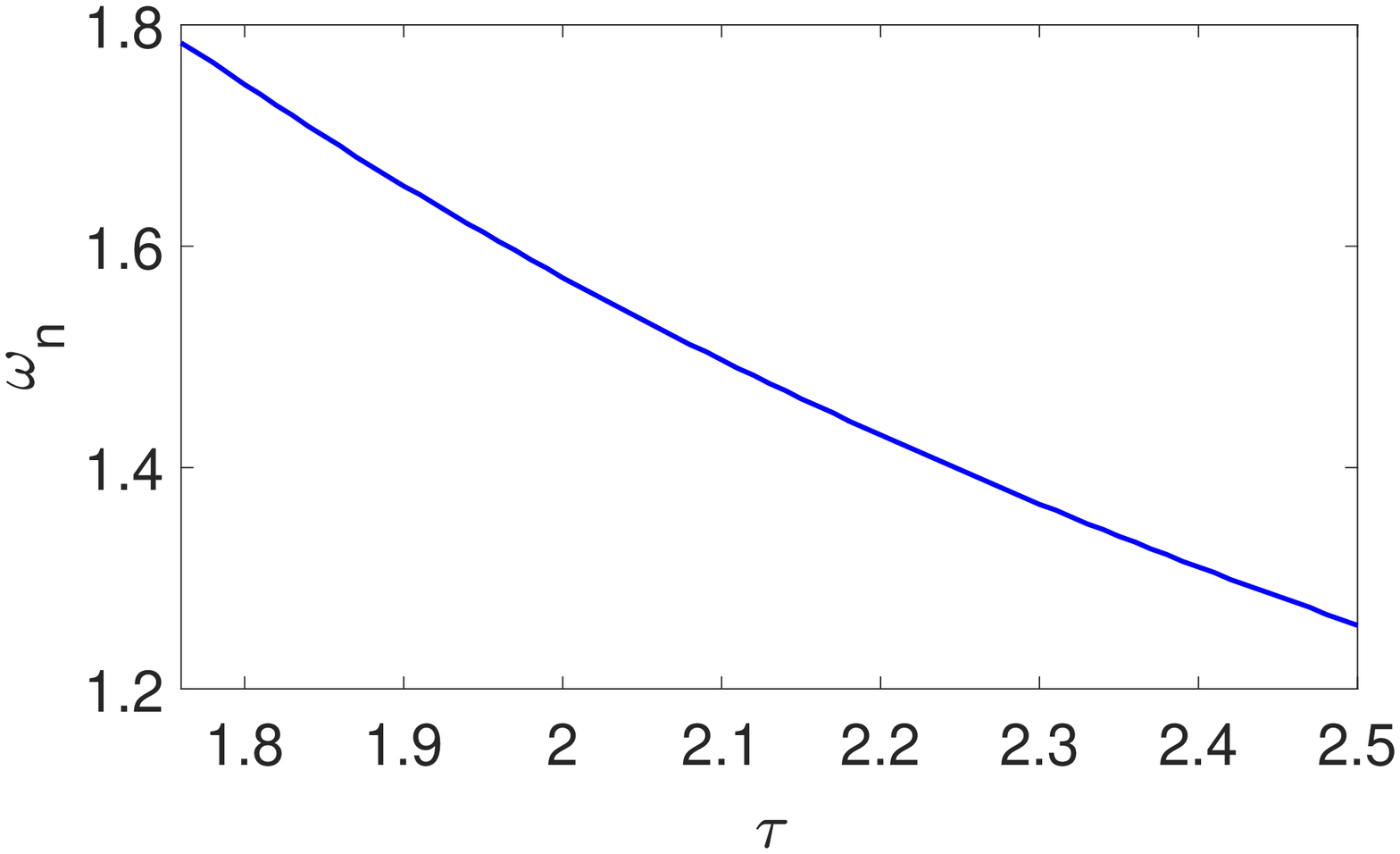} 
	\caption{\textbf{Natural frequency, $\omega_{n}$, induced by the time-delay term in function $\tau$ for $ \ddot{x}-x+0.1 x^{3} - 0.3 x (t-\tau)=0$}. For the values of $ \tau $ in the periodic range of Region II, the frequency induced by the time-delay decays monotonically with $ \tau $. }
	\centering
	\label{biflag}
\end{figure}

For the region of interest, $ \tau \in [1.76, 2.5) $, we calculate the frequency of the oscillations induced by the time-delay, that is, the natural frequency ($ \omega_{n}) $. As for the overdamped case, to find the $ \omega_{n} $ for each $ \tau $, we perform a Fourier transform to get the frequency spectrum and we name the frequency for the main peak as $ \omega_{n} $. In Fig.~\ref{biflag} we can see how $ \omega_{n} $ decreases smoothly with $ \tau $.

\subsection{Delay-induced resonance}
Once the dynamics in the unforced system has been characterized, we study the phenomenon of delay-induced resonance in the following underdamped oscillator with a periodic forcing
\begin{equation}
\ddot{x}-x+ 0.1 x^{3} - 0.3 x (t-\tau)=g\cos{\Omega t}.
\label{system}
\end{equation}

We consider the effect of the time-delay and the effect of the forcing separately in the following subsections. In the first case, the forcing produces the oscillations and we explore the possibility of using the time-delay as an enhancing term for certain values of $ \tau $. In the second case, it is the time-delay which induces the oscillations and we consider the forcing as the enhancing term for certain values of $ g $ and $ \Omega $. 

\subsubsection{Effect of the time-delay}
First, we fix the parameters of the forcing to be $ g=0.01 $ and $ \Omega=1.571 $ in Eq.~\ref{system}. Then, we consider the feedback time-delay term, $ -0.3 x(t-\tau) $, in order to enhance the system's response to this small forcing. As for the overdamped oscillator, we consider the time-delay to act as the enhancing factor if the condition $ Q(\Omega)>g $ is accomplished for some value of $ \tau $. Besides, we recall that the time-delay induces oscillations of frequency $ \omega_{n} $, which depends on $  \tau $ (Fig.~\ref{biflag}).

\begin{figure}
	\centering
	\includegraphics[width=0.75\textwidth]{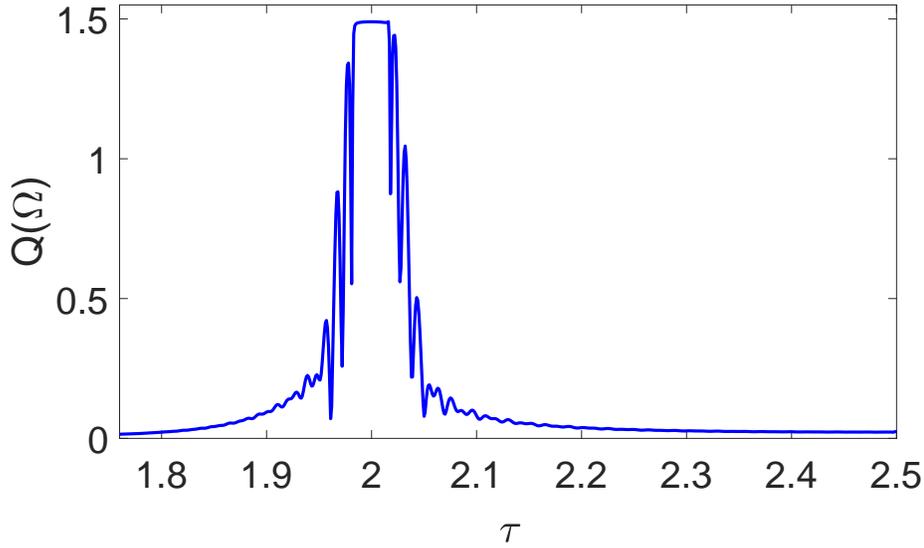} 
	\caption{\textbf{Resonance in the Q factor for the value of $ \tau $ for which $\omega_{n}=\Omega $}. The figure shows that adding a feedback time-delay term $ -0.3 x(t-\tau) $ to $ \ddot{x}-x+ 0.1 x^{3}=0.01 \cos{1.571t} $ enhances the response amplitude at the forcing frequency for a small interval around $\tau= 2$. For $\tau= 2$, the frequency induced by the time-delay is precisely $\omega_{n}=\Omega=1.571 $. }
	\label{Q_tau_underdamped_g001}
\end{figure}

Figure~\ref{Q_tau_underdamped_g001} shows the amplitude response at the forcing frequency, $ Q(\Omega) $, for values of $ \tau  $ corresponding to the periodic range of Region II. The resonance appears for an interval around $ \tau=2 $, which corresponds to a time-delay that induces oscillations with a frequency $ \omega_{n}=\Omega $. For this value of $ \tau $, the signal is enhanced as $ Q(\Omega)>g $. 

We conclude that, as for the overdamped system, the response for a driven oscillator can be enhanced by means of a time-delay that induces oscillations of the same frequency of the forcing. 

\subsubsection{Effect of the forcing}
Now, we study the conjugate phenomenon of the previous one. This time, we consider initially Eq.~\ref{system} without the forcing, which for the considered range of $ \tau $, it oscillates with frequency $ \omega_{n}$ and whose amplitude we are interested in enhancing. For that purpose, the periodic perturbation, $ g \cos{\Omega t} $, is added and we study the effect of $ g $ and $ \Omega $ on $ Q(\omega_{n}) $. As for the overdamped oscillator, if $ Q(\omega_{n})>A_{n} $, where $ A_{n} $ is the amplitude of the unforced system at the frequency $ \omega_{n} $, the signal is enhanced. 

\begin{figure}
	\centering
	\includegraphics[width=0.75\textwidth]{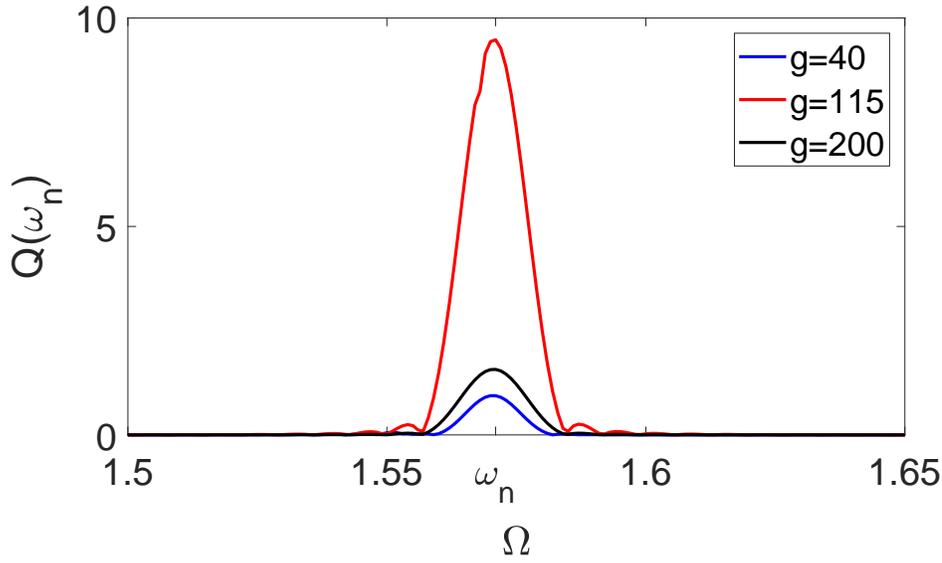} 
	\centering
	\caption{\textbf{Resonance in the Q factor for a forcing with frequency $ \Omega=\omega_{n} $}. Resonance in $ \ddot{x}-x+ 0.1 x^{3} - 0.3 x (t-2)=g\cos{\Omega t} $ appears when the frequency $ \Omega $ equals the frequency of the oscillations induced by the time-delay term independently of the value of $ g $. The amplitude of the resonance peak grows non-monotonically with $ g $.}
	\label{Qvariosg}
\end{figure}

Figure~\ref{Qvariosg} shows that a resonance appears for $ \Omega=\omega_{n} $  independently of the value of $ g $. For this frequency, the response amplitude at the frequency $ \omega_{n} $ is greatly enlarged. In fact, the system is no longer confined to one of the wells. This means that when a forcing with the same frequency of the intra-well oscillations ($ \Omega= \omega_{n} $) is added, we can make the system oscillate in both wells with the same frequency that it oscillated in one of them. As for the overdamped oscillator, this is a conjugate phenomenon: it is possible to enhance the response of a driven oscillator by means of a time-delay with $ \omega_{n}=\Omega $ and also it is possible to enhance the oscillations induced by a time-delay by means of a periodic forcing with $ \Omega=\omega_{n} $. 

It should be noted that the peak's amplitude does not always grow with $ g $. We can see how it reaches its biggest amplitude for $ g $=115, while for $ g $=200 and $ g $=40 the amplitude is significantly smaller. This is an important difference between the underdamped and the overdamped case, where the amplitude grows monotonically with $ g $. 

\begin{figure}
	\centering
	\includegraphics[width=0.75\textwidth]{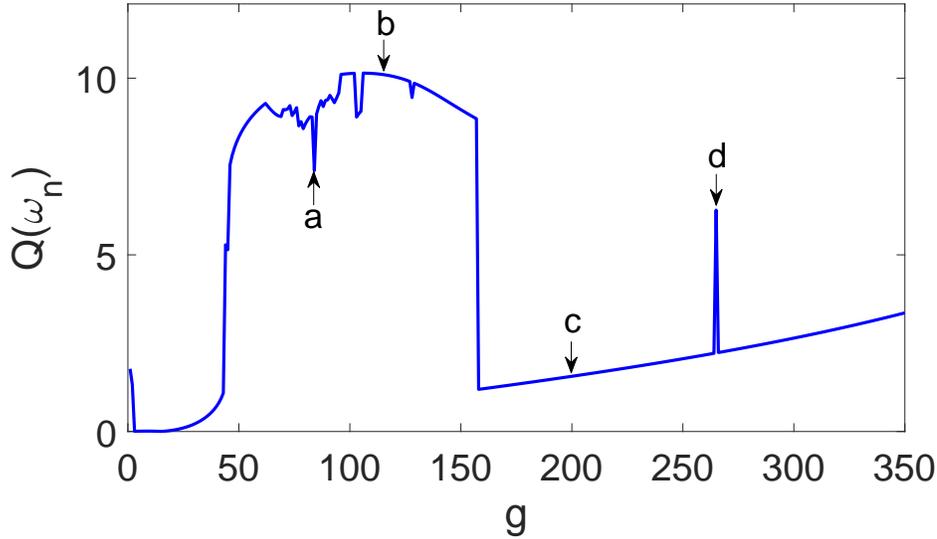} 
	\centering
	\caption{\textbf{Resonance in the Q factor for an interval of values of the amplitude of the forcing, $ g $}. The figure shows explicitly the dependence of the Q factor for $ \Omega=\omega_{n}=1.571 $ and a time-delay term $ -0.3 x(t-2) $ with the amplitude of the forcing. For  $ g \in [44, 157]$ the Q factor is greatly enhanced presenting resonant behavior. The arrows point at some relevant values: $g=84, 115, 200$ and $265$.}
	\label{w1coma633}
\end{figure}

In Fig.~\ref{w1coma633} we show explicitly the effect of $ g $ on the Q factor. It is observed that for $ g \in [44, 157]$ the Q factor is greatly enhanced. After this region, for bigger values of $ g $, the Q factor falls again. Its main implication is that it is not only important to tune the frequency of the forcing but also its amplitude if we want to enhance the unforced system's oscillations. About the effect of $ g $, the most important limitation, as in VR \cite{Landa2000}, lies in the fact that the forcing needed for resonance is not \textit{small} compared to the amplitude of the unforced oscillator. However, the response amplitude at $ \omega_{n} $ is greatly enhanced for a wide range of values of $ g $. 

\begin{figure}
	\begin{center}
		\includegraphics[width=7cm]{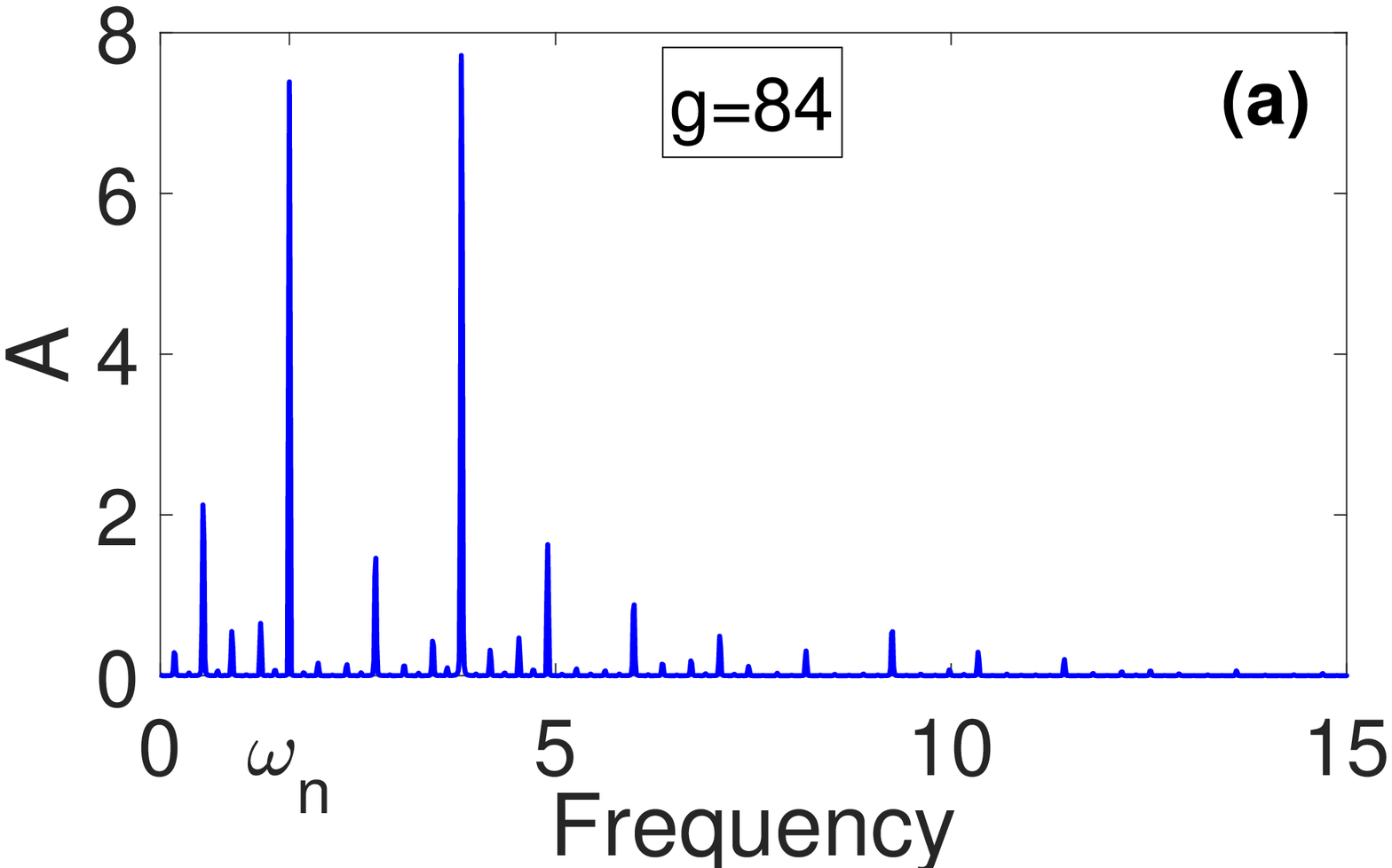}
		\includegraphics[width=7cm]{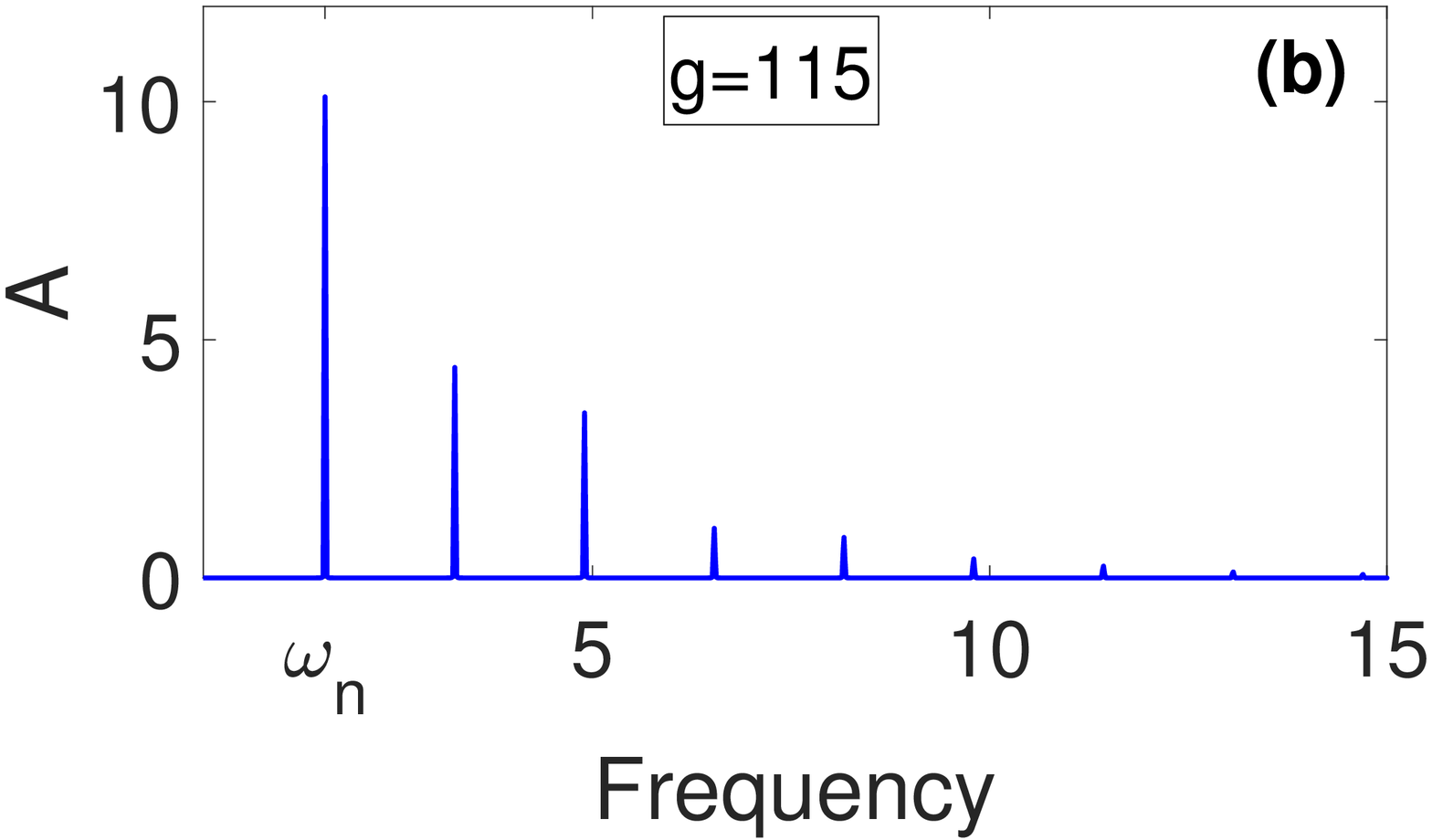} \\
		\includegraphics[width=7cm]{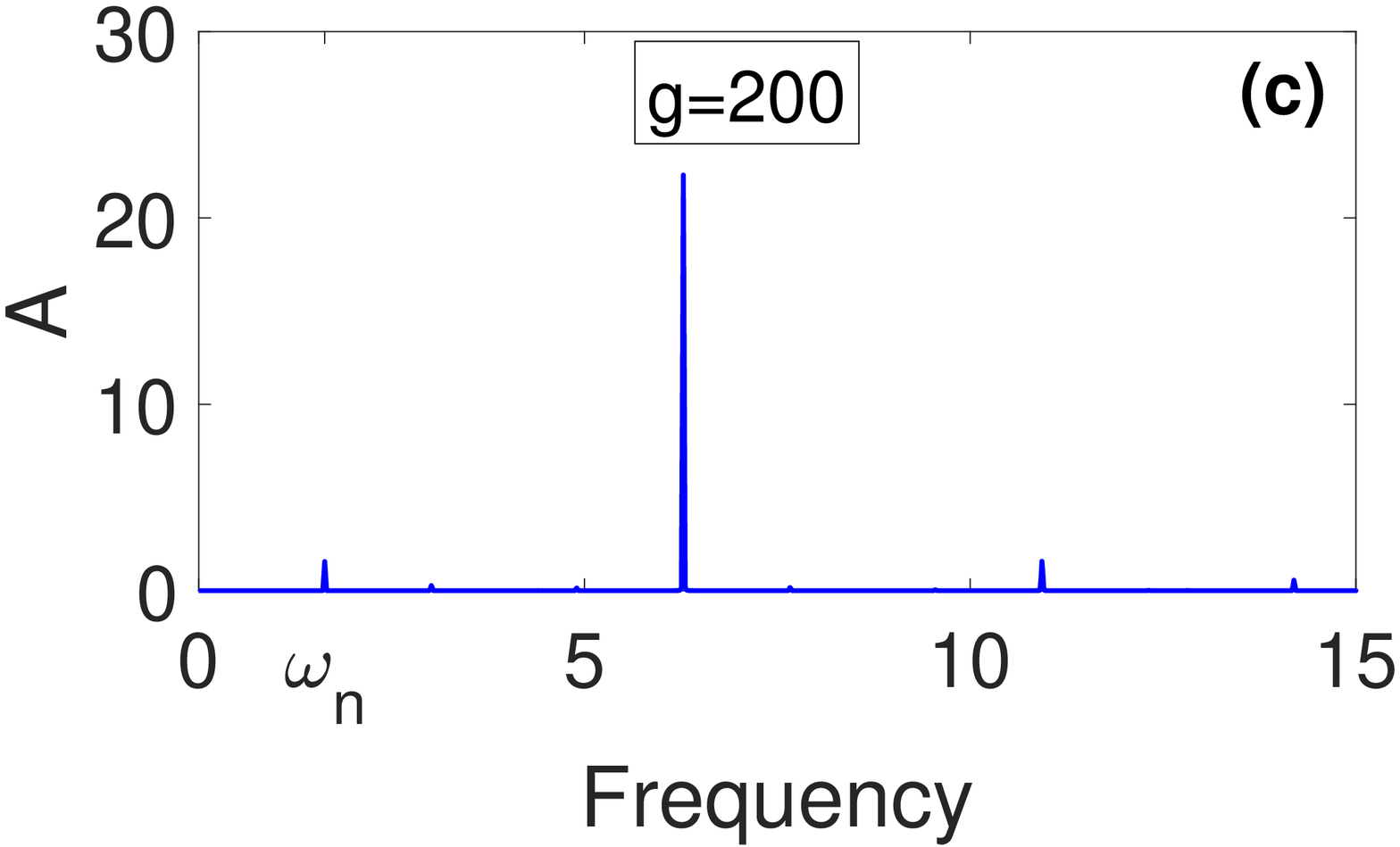}
		\includegraphics[width=7cm]{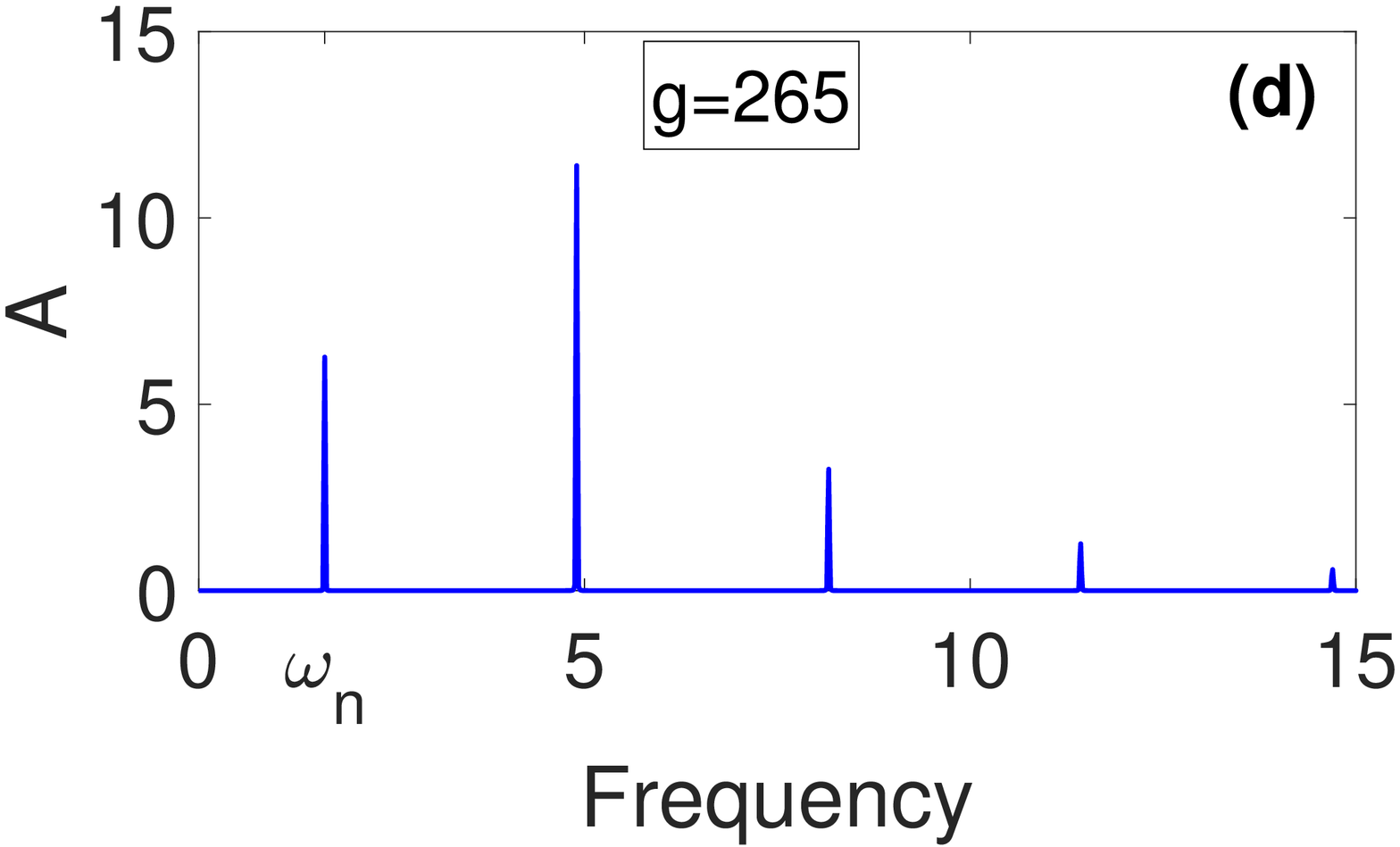}
	\end{center}
	\caption{\textbf{Frequency spectrum for the response of the underdamped oscillator $\ddot{x}-x+ 0.1 x^{3} - 0.3 x (t-2)=g\cos{1.571 t} $ for different values of $ g $.} The frequency spectra were calculated for a frequency $ \Omega=\omega_{n}=1.571 $ and for some different values of $ g $ to show how the forcing amplitude influences the response frequency. Panel (b) corresponding to $ g =115$ belongs to the resonant and smooth region in Fig.~\ref{w1coma633} and it is the only one for which the peak for $ \omega_{n} $ in the spectrum is the main peak. For values of $ g $ that do not belong to the resonance region (Panels (a)-(c)-(d)), other frequencies have a bigger amplitude. This implies that the forcing does not serve as an effective enhancer for the oscillations induced by the delay.}
	\label{FFT}
\end{figure}

For certain values of $ g $, the curve for the Q factor in Fig.~\ref{w1coma633} looses smoothness and presents some peaks. We have studied in detail the system's behavior at the points named a, b, c and d for a better understanding. For that purpose the frequency spectrum of the system's response at these points is presented in Fig.~\ref{FFT}.

Figures~\ref{FFT}(a)-(d) correspond to singular values of $ g $. The value of $ g=84 $ (point a), belongs to the non-smooth region and in the spectrum this is reflected by the presence of numerous peaks. In fact, it was observed that the non-smooth region present in Fig.~\ref{w1coma633} corresponds to the parameters for which the system's behavior is aperiodic. 

Figure~\ref{FFT}(b) shows the frequency spectrum for $ g =115$ (point b), which corresponds to the top of the resonant region. In this spectrum a main peak appears for $ \omega=1.571 $, which is precisely what we called the natural frequency of the system without forcing.

For $ g = 200$ (point c), the Q factor falls from the resonance region and this is reflected in the frequency spectrum in Fig.~\ref{FFT}(c) as the peak for $ \omega=1.571 $ is much more smaller than in the previous case and is approximately of the same height that without the forcing. 

The case of $ g=265 $ (point d), is an exception as it corresponds to a peak in the smooth region after the resonance. As we can see in the spectrum, Fig.~\ref{FFT}(d), the peak's amplitude for $ \omega=1.571 $ is bigger than in the rest of the region after the resonance (compare it to $ g=200 $) although it is not the main peak as in the resonance region.

\section{Discussion and conclusions} \label{Sdiscussion}
In the present work we have focused on exploring the delay-induced resonance for the forced and time-delayed Duffing oscillator considering both the overdamped and underdamped cases. 

In the case of the overdamped oscillator $ \dot{x} - x +  x^{3} -\gamma x (t-\tau) =g \cos{\Omega t} $, we have performed a numerical and analytical study of the bifurcations. Then, for the parameter values for which the time-delay induces oscillations of frequency $ \omega_{n}(\gamma, \tau) $, we have shown that the oscillations caused by the forcing may be enhanced by means of a time-delay term with $ \omega_{n}=\Omega $. Furthermore, we have shown that the conjugate phenomenon is also possible, that is, the oscillations induced by the time-delay may be enhanced by means of a forcing with frequency $ \Omega=\omega_{n} $. This is an interesting result since the nature of both perturbations is different.

On the other hand, we have considered the parameter values for which the time-delay does not induce oscillations in the steady state. Even for these values, the time-delay still induces a resonance: the amplitude of the oscillations produced by the forcing depends on the frequency of the forcing, $ \Omega $, and $ \tau $, presenting a resonant behavior. 

Turning now to the underdamped oscillator $ \ddot{x}+\omega_{0}^{2}x+\beta x^{3} +\gamma x (t-\tau)=g\cos{\Omega t} $, we have performed a numerical analysis of the bifurcations, and we have focused on the range of values of $ \tau $ that induce periodic oscillations confined to one well with frequency $ \omega_{n}(\gamma, \tau)$. As for the overdamped oscillator, we have shown that the response can be enhanced by means of the time-delay as well as the oscillations induced by the time-delay can be enhanced by means of the forcing when the frequencies $ \omega_{n}=\Omega $. As a difference, the Q factor varies non-monotonically with the amplitude of the forcing, $ g $, in the underdamped case.

\nonumsection{Acknowledgments} \noindent 
This work has been supported by the Spanish State Research Agency (AEI) and the European Regional Development Fund (FEDER) under Project No. FIS2016-76883-P.

\bibliography{Cacosesa_ijbc_Revised_August_19}
\bibliographystyle{ws-ijbc}

\end{document}